\documentstyle[12pt,epsf,epsfig,a4]{article}
\begin{document}
\begin {titlepage}
\begin{flushleft}
FSUJ TPI QO-2/97
\end{flushleft}
\begin{flushright}
March, 1997
\end{flushright}
\vspace{20mm}
\begin{center}
{\Large {\bf Photon-added state preparation via conditional measurement on a 
beam splitter} \\[3ex]
\large M. Dakna, L. Kn\"oll, D.--G. Welsch}\\[2.ex]
Friedrich-Schiller-Universit\"at Jena 
Theoretisch-Physikalisches Institut \\[1ex]
Max-Wien Platz 1, D-07743 Jena, Germany
\vspace{25mm}
\end{center}
\begin{center}
\bf{Abstract}
\end{center}
We show that conditional output measurement on a beam splitter may
be used to produce photon-added states for a large class of
signal-mode quantum states, such as thermal states, coherent states, 
squeezed states, displaced photon-number states, and coherent 
phase states. Combining a mode prepared in such a state and a mode 
prepared in a photon-number state, the state of the mode in one 
of the output channels of the beam splitter ``collapses'' to a photon-added 
state, provided that no photons are detected in the other output channel.
We present analytical and numerical results, with special emphasis
on photon-added coherent and squeezed vacuum states. In particular, we show 
that adding photons to a squeezed vacuum yields superpositions of quantum
states which show all the typical features of Schr\"{o}dinger-cat-like
states.
\end{titlepage}
\renewcommand {\thepage} {\arabic{page}}
\setcounter {page} {2}
\section{Introduction}
\label{sec1}
It is well known that according to von Neumann's projection principle
\cite{vNeumann1} conditional measurement may be a fruitful method
for quantum-state manipulation and engineering. In particular,
when a system, such as a correlated two-mode optical field or a
correlated atom-field system, is prepared in an entangled state of
two subsystems and a measurement is performed on one subsystem, then
the quantum state of the other subsystem can be reduced to a new state.
Systems that have typically been considered are the waves produced by
parametric amplifiers \cite{Watanabe1,Song,Ban1} and degenerate
four-wave mixers \cite{Ban2}, the interfering fields in the output
channels of a beam splitter \cite{Ban1,Ban3}, and systems of the
Jaynes-Cummings type in cavity QED \cite{Brune1} or trapped-ion studies
\cite{Cirac1}. Further, state reduction via continuous measurement
has also been considered \cite{Ueda1,Ueda2,Ogawa1,Ban1}.
Conditional measurement offers new possibilities of generating extremely
nonclassical states, such as photon-number states
\cite{Ban2,Cirac1,Ueda2,Milburn1,Paul2} and  Schr\"odinger-cat-like
states \cite{Song,Ogawa1,Dakna1}.

In this paper we show that zero-photon conditional output measurement 
on a beam splitter can be used advantageously to generate photon-added
states for a large class of quantum states of the signal mode,
such as thermal states, coherent states, squeezed states, displaced 
photon-number states, coherent phase states etc.. It can be expected 
that repeated application of the photon creation
operator to the signal-mode quantum state can produce extremely
nonclassical states. Photon adding can therefore be expected to improve
the performance of noise reduction schemes \cite{Gunnar1}.
In particular, the states that are obtained from
coherent states by repeatedly applying to them the photon creation
operator -- the photon-added coherent states -- 
can be regarded as non-Gaussian squeezed states and were first introduced
and discussed in \cite{Agarwal1} and it was proposed that they can
be produced in nonlinear processes in cavities. 
We show that photon-added coherent states can also be generated
via conditional output measurement on a beam splitter,
mixing a signal mode prepared in a coherent state with a second mode
prepared in a photon-number state.

We further show that when a signal mode prepared in a squeezed vacuum is 
mixed with photon-number states and zero-photon conditional
output measurements are performed, then photon-added squeezed vacuum states
can be produced. We analyze the states in terms of the photon-number
and quadrature-component distributions and the Wigner and
Husimi functions. We show that photon-added squeezed vacuum states  
exhibit all the typical properties of Schr\"{o}\-din\-ger-cat-like
states, so that photon adding can be regarded as a method for producing
Schr\"{o}dinger cats. In particular, they are shown to be superpositions 
of two non-Gaussian squeezed coherent states that tend to Gaussian squeezed 
coherent states for sufficiently large number of added photons. 
It is worth noting that although the scheme bears some resemblance to 
that in \cite{Dakna1}, the two schemes are quite different from each other. 
That concerns both the second input quantum state (which is the vacuum 
in \cite{Dakna1}) and the conditional output measurement (detection 
of a nonzero number of photons in \cite{Dakna1}).

The paper is organized as follows. In Sec.~\ref{sec2} the basic scheme
is explained and the conditional output states are derived. The
possibility of the generation of photon-added states is studied
in Sec.~\ref{sec3}, with special emphasis on photon-added coherent and
squeezed vacuum states. The problem of mixed photon-added states
is addressed in Sec.~\ref{sec4}. A summary and concluding remarks 
are given in Sec.~\ref{sec5}.


\section{Basis equations}
\label{sec2}

It is well known that the input--output relations at a lossless beam
splitter can be characterized by the $\rm SU(2)$ Lie algebra
\cite{Yurke3,Campos1}. In the Heisenberg picture, the photon destruction
operators of the outgoing modes, $\hat{b}_{k}$ ($k\!=\!1,2$), can be
obtained from those of the incoming modes, $\hat{a}_{k}$, as
\begin{equation}
\hat{b}_{k} = \sum_{k'=1}^{2} T_{k,k'} \, \hat{a}_{k'},
\label{1.01}
\end{equation}
where
\begin{eqnarray}
(T_{k,k'}) = e^{i\varphi_0}
\left(\begin{array}{cc}\cos\theta
\;e^{i\varphi_{T}}&\sin\theta\;e^{i\varphi_{R}}\\
-\sin\theta\;e^{-i\varphi_{R}}&\cos\theta\;e^{-i\varphi_{T}}
\end{array}\right)
\label{1.02}
\end{eqnarray}
is a SU(2) matrix whose elements are given by the complex
transmittance $T$ and reflectance $R$ of the beam splitter,
\begin{equation}
\label{1.02a}
T = \cos\theta\;e^{i\varphi_{T}}, \quad 
R = \sin\theta\;e^{i\varphi_{R}}. 
\end{equation}
Using the Schr\"{o}dinger picture, the photonic operators
are left unchanged, but the density operator is transformed. In this
case the output-state density operator $\hat{\varrho}_{\rm out}$ can be
related to the input-state density operator $\hat{\varrho}_{\rm in}$ as
\begin{equation}
\hat{\varrho}_{\rm out}
= \hat{V}^{\dagger} \hat \varrho_{\rm in}  \hat{V},
\label{1.03}
\end{equation}
where $\hat{V}$ can be given by \cite{Yurke3,Campos1}
\begin{equation}
\hat{V} =
e^{-i(\varphi_{T}-\varphi_{R}) \hat{L}_{3}}
\, e^{-2i\theta \hat{L}_{2}}
\, e^{-i(\varphi_{T}+\varphi_{R}) \hat{L}_{3}},
\label{1.03a}
\end{equation}
with
\begin{equation}
\hat{L}_{2} = \textstyle\frac{1}{2i}(\hat{a}_{1}^\dagger\hat{a}_{2}
-\hat{a}_{2}^\dagger\hat{a}_{1}), \quad
\hat{L}_{3} = \textstyle\frac{1}{2}(\hat{a}_{1}^\dagger\hat{a}_{1}
-\hat{a}_{2}^\dagger\hat{a}_{2}).
\label{1.03b}
\end{equation}
Note that $\varphi_{0}$ is a global phase factor that may be omitted
without loss of generality, $\varphi_{0}\!=\!0$. Applying
elementary parameter-differentiation techniques \cite{Wilcox},
we can derive the operator identity
\begin{equation}
e^{-2i\theta \hat{L}_{2}}=e^{\tan(\theta) \hat{a}_{2}^\dagger\hat{a}_{1}}
\,e^{2\ln\cos(\theta)\hat{L}_3}
\,e^{-\tan(\theta) \hat{a}_{1}^\dagger\hat{a}_{2}},
\end{equation}
which [together with Eq.~(\ref{1.02a})] enables us to
rewrite $\hat{V}^{\dagger}$, Eq.~(\ref{1.03a}), as
\begin{equation}
\hat{V}^{\dagger} =
T^{\hat{n}_{1}}
\, e^{-R^*\hat{a}^\dagger_{2}\hat{a}_{1}}
\, e^{R\hat{a}^\dagger_{1}\hat{a}_{2}}
\,T^{-\hat{n}_{2}}\,,
\label{1.03c}
\end{equation}
where  $\hat{n}_{k}=\hat{a}^\dagger_{k}\hat{a}_{k}$.

\begin{figure}
\centering\epsfig{figure=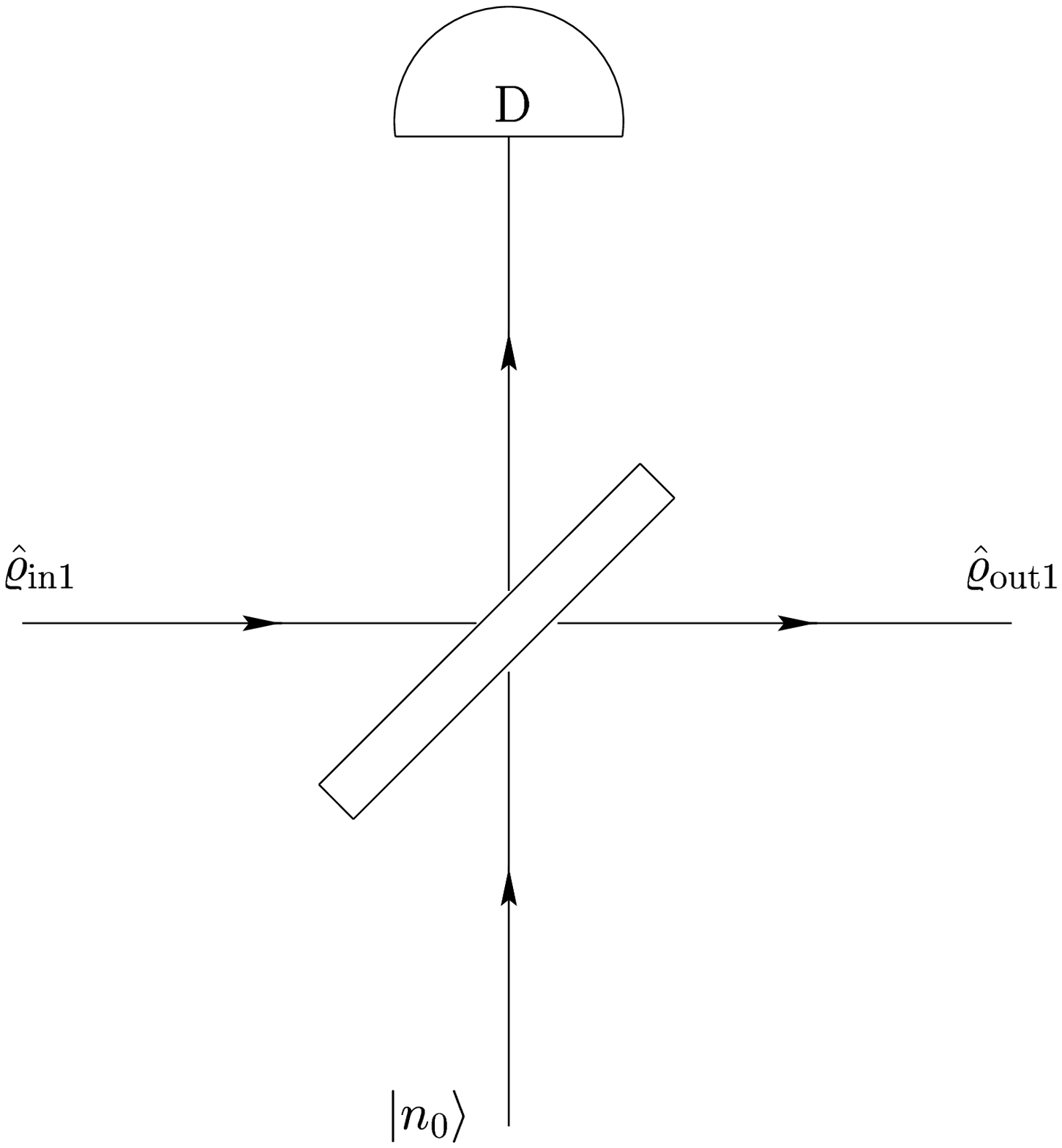,width=0.9\linewidth}
\caption{Scheme of the experimental setup. When a signal
mode prepared in a state $\hat{\varrho}_{{\rm in}1}$
is mixed with another input mode prepared in a Fock state $|n_{0}\rangle$
and in one of the output channels of the beam splitter no photons are
recorded by a detector (D), then the quantum state 
$\hat{\varrho}_{{\rm out}1}$ of the mode in the other output channel 
``collapses'' to a photon-added state.
\label{Fig1}}
\end{figure}

Now, let us assume (Fig.~\ref{Fig1}) that the modes 
that are fed into the first and second input channels of the 
beam splitter are prepared in a state described by a density 
operator $\hat{\varrho}_{{\rm in}1}$ and a Fock state 
$\hat{\varrho}_{{\rm in}2}$ $\!=$ $\!|n_{0}\rangle\langle n_{0}|$, 
respectively (for a review on the problem of the generation of Fock 
states, see \cite{Davidovich1}). The input-state density operator can 
then be written as
\begin{equation}
\hat \varrho_{\rm in}(n_{0}) = \hat \varrho_{{\rm in}1}
\otimes |n_0\rangle \langle\,n_0|.
\label{1.04}
\end{equation}
Using Eqs.~(\ref{1.03}), (\ref{1.03c}), and (\ref{1.04}),
after some calculation the output-state density operator
$\hat{\varrho}_{\rm out}$ $\!\equiv$ $\!\hat{\varrho}_{\rm out}(n_{0})$ 
can be given by
\begin{eqnarray}
\lefteqn{
\hat \varrho _{\rm out}(n_0) = \frac{1}{|T|^{2n_0}}
\sum_{n_{2}=0}^{\infty}\sum_{m_{2}=0}^{\infty}\sum_{k=0}^{n_0}
\sum_{j=0}^{n_0} (R^*)^{m_2+j} R^{n_2+k} 
}
\nonumber \\  &&  \times 
\frac{(-1)^{n_2+m_2} }{ \sqrt{k!j!m_2!n_2!} }
\sqrt{\! {n_0\choose k}\!{n_0\choose j}\!
{n_0\!-\!k\!+\!m_2\choose m_2}\!{n_0\!-\!j\!+\!n_2\choose n_2} }  
\nonumber \\  &&  \times \,
T^{\hat{n}_{1}}
{\hat{a}_{1}}^{m_2}({\hat{a}_{1}^\dagger})^{k}
\hat{\varrho}_{{\rm in}1}\hat{a}_{1}^{j}
({\hat{a}_{1}^\dagger})^{n_{2}}(T^*)^{\hat{n}_{1}}
\otimes 
|n_0-k+m_{2}\rangle\langle n_0-j+n_{2}|.
\label{1.0}
\end{eqnarray}
It can easily be seen that when the second input channel is unused, i.e., $n_0$ $\!=$
$\!0$, then Eq.~(\ref{1.0}) reduces to the relation considered in
\cite{Ban1,Ban3,Dakna1}.

  From Eq.~(\ref{1.0}) we see that the output modes are, in general,
highly correlated. When the photon number of the mode in the second
output channel is measured and $m_2$ photons are detected, then the mode
in the first output channel is prepared in a quantum state whose
density operator $\hat{\varrho}_{{\rm out}1}(n_{0},m_{2})$ reads as
\begin{eqnarray}
\hat{\varrho}_{{\rm out}1}(n_0,m_2)
= \frac{\langle m_{2} |\hat{\varrho}_{\rm out}(n_0) | m_{2} \rangle}
{{\rm Tr}_{1}( \langle m_{2} |
\hat{\varrho}_{\rm out}(n_{0}) | m_{2} \rangle)} \,.
\label{1.1}
\end{eqnarray}
The probability of such an event is given by
\begin{eqnarray}
\lefteqn{
\hspace*{-2ex}
P(n_0,m_2)=
{\rm Tr}_{1}( \langle m_{2} |
\hat{\varrho}_{\rm out}{(n_0)} | m_{2} \rangle )
= \!\! \sum_{n_{1}=\mu-\nu}^{\infty}
\sum_{j=\mu}^{n_0}\sum_{k=\mu}^{n_0}\;
\!\!|R|^{2(j+k-\nu)}|T|^{2(n_1+\nu-n_0)}
}
\nonumber \\ && \hspace{-4ex}\times \,
\frac{(-1)^{j+k}n_0!\,n{_1}!}{(n_1+\nu)!(n_0-\nu)!}
{n_0-\nu\choose j-\nu} {n_0-\nu\choose k-\nu}
{n_1+j\choose j}{n_1+k\choose k}
\langle n_{1} | \hat{\varrho}_{{\rm in}1} | n_{1} \rangle,
\label{1.6}
\end{eqnarray}
where
\begin{equation}
\nu=n_0-m_2,\;\mu=\max(0,\nu).
\end{equation}


\section{Generation of photon-added states}
\label{sec3}

Let us now assume that the (signal) mode in the first input channel is prepared 
in a mixed state 
\begin{equation}
\hat{\varrho}_{{\rm in}1}
= \sum_{\Phi} \tilde p_{\Phi} \, | \Phi\rangle\langle \Phi  |
\label{1.61}
\end{equation}
($\sum_{\Phi}\tilde p_{\Phi}$ $\!=$ $\!1$, $0$ $\!\leq$ $\!\tilde p_{\Phi}$
$\!\leq$ $\!1$) and restrict attention to the events 
that no photons are recorded in the second output channel, i.e.,
\begin{eqnarray}
\label{1.61aa}
m_{2} = 0, \qquad \nu = n_{0}
\end{eqnarray}
in Eqs.~(\ref{1.0}) -- (\ref{1.6}).
Note that such events can be detected using highly efficient avalanche
photodiodes. Combining Eqs.~(\ref{1.0}) and (\ref{1.1}) and using
Eqs.~(\ref{1.61}) and (\ref{1.61aa}), we find that the
mode in the first output channel is prepared in a state
\begin{equation}
\hat{\varrho}_{{\rm out}1}(n_0,m_{2}=0)
= \sum_{\Phi} \tilde p_{\Phi}| \, \Psi_{n_0} \rangle \big\langle \Psi_ {n_0}| \,,
\label{1.3}
\end{equation}
where
\begin{eqnarray}
|\Psi_{n_0}\rangle=
\frac{1}{\sqrt{{\cal N}_{n_0}}}
\, (\hat{a}^\dagger_1)^{n_0}\,T^{\hat{n}_1} |\Phi\rangle ,
\label{1.4m}
\end{eqnarray}
${\cal N}_{n_{0}}$ being a normalization constant,
\begin{eqnarray}
\label{1.4ma}
{\cal N}_{n_{0}}
= \langle \Phi | (T^{\ast})^{\hat{n}_1} \hat{a}_{1}^{n_0}
(\hat{a}^\dagger_1)^{n_0} T^{\hat{n}_1} | \Phi \rangle .
\end{eqnarray}
The probability of observing the conditional state 
$\hat{\varrho}_{{\rm out}1}(n_0,m_{2}=0)$ can easily be 
found from Eq.~(\ref{1.6}) and reads
\begin{eqnarray}
P(n_{0}) \equiv P(n_0,m_2=0)
|R|^{2n_0}\sum_{n_1=0}^{\infty}
|T|^{2n_1}
{n_1+n_0\choose n_0 }
\langle n_{1} | \hat{\varrho}_{{\rm in}1}| n_{1} \rangle.
\label{M1}
\end{eqnarray}

The states $|\Psi_{n_{0}}\rangle$, Eq.~(\ref{1.4m}), are obviously the 
conditional states observed in the case when the mode in the first 
input channel is prepared in a pure state $|\Phi\rangle$. From 
Eq.~(\ref{1.4m}) we see that for chosen $n_{0}$ the conditional
state $|\Psi_{n_{0}}\rangle$ is a photon-added state, with $n_{0}$
photons being added to the state $|\Psi\rangle$ $\!\sim$
$\!T^{\hat n_{1}} |\Phi\rangle$. In particular when the absolute value
of the transmittance is close to unity, $|T|$ $\!\approx$ $\!1$, then the
state $|\Psi\rangle$ is -- apart from a rotation in the phase space -- close
to the input state $|\Phi\rangle$. To be more specific, let us consider
the expansion of $|\Phi\rangle$ in the Fock basis,
\begin{eqnarray}
\label{1.62}
| \Phi\rangle=\sum_{n_1=0}^{\infty}c_{n_1}|n_1\rangle ,
\end{eqnarray}
and assume that
\begin{eqnarray}
\label{1.62a}
c_{n_1} \approx 0  \quad {\rm for} \quad  n_{1} > n_{\rm max}.
\end{eqnarray}
We see that (apart from the rotation mentioned) $|\Psi\rangle$
$\!\approx$ $|\Phi\rangle$, provided that
\begin{eqnarray}
\label{1.62b}
|T|^{n_{1}} \approx 1  \quad {\rm for} \quad n_{1} \leq n_{\rm max}.
\end{eqnarray}
Note that for any physical state $|\Phi\rangle$ the expansion in
Eq.~(\ref{1.62}) can always be approximated to any desired degree of
accuracy by truncating it at $n_{\rm max}$ if $n_{\rm max}$ is
suitably large.

However, there are classes of states for which the corresponding
photon-added states can be produced even when $|T|$ is not close
to unity. Let us consider a class of
$\alpha$ parametrized states $|\Phi(\alpha)\rangle$ such that
\begin{eqnarray}
\label{1.62c}
| \Phi(\alpha)\rangle=\sum_{n_1=0}^{\infty}c_{n_1}(\alpha)|n_1\rangle ,
\quad {\rm with} \quad c_{n_1}(\alpha) \propto \alpha^{n_{1}}.
\end{eqnarray}
Since in this case the relation $|\Psi\rangle$ $\!\sim$ $\!T^{\hat n_{1}}
|\Phi(\alpha)\rangle$ $\!=$ $\!|\Phi(T\alpha)\rangle$ is valid,
the state $|\Psi\rangle$
obviously belongs to the class of states $|\Phi(\alpha)\rangle$.
Hence when the input state $|\Phi\rangle$ belongs to the class of states
$|\Phi(\alpha)\rangle$, then the conditional output state
\begin{eqnarray}
\label{1.62d}
|\Psi_{n_{0}}\rangle \sim (\hat{a}^{\dagger})^{n_{0}} |\Psi\rangle
\sim (\hat{a}^{\dagger})^{n_{0}} |\Phi(T\alpha)\rangle
\end{eqnarray}
is a photon-added states, with the photons being added to a state
$|\Psi\rangle$ that also belongs to the class of states
$|\Phi(\alpha)\rangle$.
The extension of the above given considerations to mixed states is
straightforward.   
Typical examples of classes of photon-added states that can
be produced in this way are thermal states, coherent states,
squeezed states, displaced Fock states, and coherent phase states.

When the number of added photons, $n_{0}$, is sufficiently small compared
to the photon numbers that mainly contribute to the state $|\Psi\rangle$
and the photon-number distribution $|\langle n_{1} |\Psi\rangle|^{2}$ is
sufficiently slowly varying with $n_{1}$, then the photon-added state
$|\Psi_{n_{0}}\rangle$ exhibits, in general, properties that are
similar to those of the state $|\Psi\rangle$. Writing
\begin{eqnarray}
|\Psi_{n_{0}}\rangle \sim (\hat{a}^{\dagger})^{n_{0}} |\Psi\rangle \sim
\sum_{n_1=0}^{\infty}\frac{\tilde{c}_{n_1}}{\tilde{c}_{n_1+n_0}}
\,\tilde{c}_{n_1+n_0}
\left[\frac{(n_1+n_0)!}{n_1!}\right]^{1/2}|n_1+n_0\rangle
\end{eqnarray}
($\tilde{c}_{n_{1}}$ $\!=$ $\!T^{n_{1}}c_{n_{1}}$) and using
the approximations
\begin{eqnarray}
\frac{\tilde{c}_{n_1}}{\tilde{c}_{n_1+n_0}}\approx 1, \quad
\left[\frac{(n_1+n_0)!}{n_1!}\right]^{1/2}\approx (\bar{n})^{n_0/2},
\end{eqnarray}
we approximately derive
\begin{eqnarray}
|\Psi_{n_{0}}\rangle \sim |\Psi\rangle \sim 
\sum_{n_1=0}^{\infty}\tilde{c}_{n_1}|n_1\rangle,
\end{eqnarray}
i.e., photon adding leaves the state $|\Psi\rangle$ nearly unchanged.
With increasing number of added photons qualitatively
new properties are expected to be observed. To illustrate the method,
let us consider photon-added coherent and squeezed vacuum states
in more detail. 


\subsection{Coherent states}
\label{sec3a}

Let us first assume that the input field is prepared in a coherent state,
i.e., $\hat{\varrho}_{{\rm in}1}$ $\!=$ $\!|\Phi\rangle\langle\Phi |$,
where
\begin{equation}
|\Phi\rangle \equiv 
|\beta\rangle = e^{-|\beta|^2/2}\sum_{n=0}^{\infty}\frac{\beta^n}
{\sqrt{n!}}\,|n\rangle,
\label{Q1}
\end{equation}
with $\beta=|\beta|e^{\varphi_\beta} $. According to Eq.~(\ref{1.4m}),
the conditional output states then reads
\begin{equation}
|\Psi_{n_0}\rangle=\frac{1}{\sqrt{{\cal N}'_{n_0}}}
(\hat{a}^\dagger)^{n_0}|\beta'\rangle,
\label{Q2}
\end{equation}
where $\beta'$ $\!=$ $\!T \beta$ and
\begin{equation}
{\cal N}'_{n_0}
= n_0!\,{\rm L}_{n_0}\!\left(-|\beta'|^2\right),
\label{Q2v}
\end{equation}
${\rm L}_{n_0}(x)$ being the Laguerre polynomial. The states
are photon-added coherent states and can be represented in the Fock 
basis as
\begin{equation}
|\Psi_{n_0}\rangle=\frac{e^{-|\beta'|^2/2}}{\sqrt{{\cal N}'_{n_0}}}
\sum_{n=0}^{\infty}\frac{(\beta')^n}{\sqrt{n!}}
\left[\frac{(n+n_0)!}{n!}\right]^{1/2}  |n+n_0\rangle.
\label{Q2w}
\end{equation}

\begin{figure}
\centering\epsfig{figure=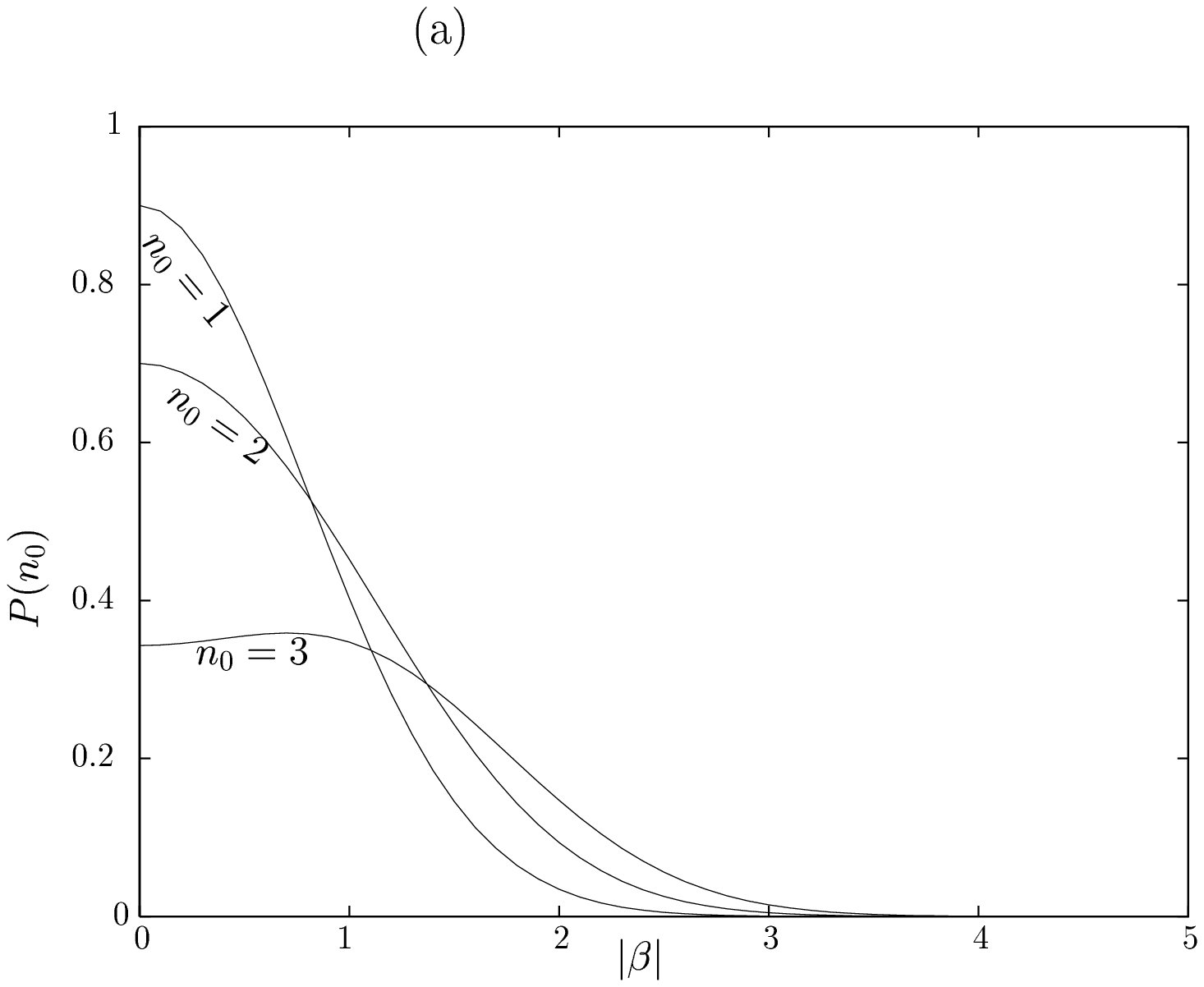,width=0.8\linewidth}

~
\vspace{0.1cm}

~

\centering\epsfig{figure=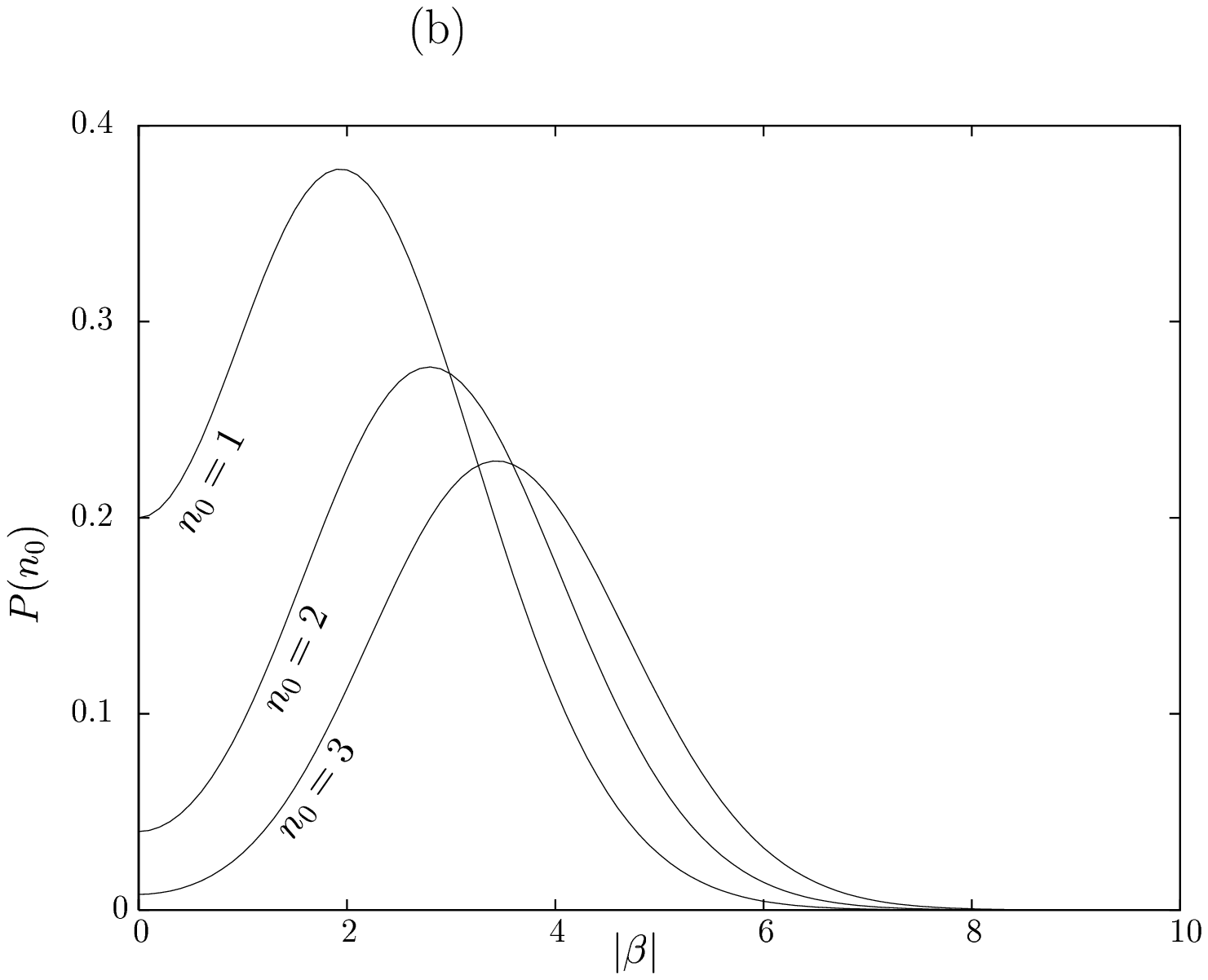,width=0.8\linewidth}
\caption{The probability of producing photon-added coherent
states is shown as a function of $|\beta|$ for two values of the
beam-splitter transmittance [(a) $|T|^2$ $\!=$ $\!0.3$;
(b) $|T|^2$ $\!=$ $\!0.8$] and various values of $n_{0}$.
\label{Fig2}}
\end{figure}

   From Eqs.~(\ref{M1}) and (\ref{Q1}), the probability of
producing photon-added coherent states is given by
\begin{eqnarray}
P(n_0)
=|R|^{2n_0}e^{-|\beta|^2}\sum_{n=0}^{\infty}{n+n_0\choose n_0 }
\frac{|\beta'|^{2n}}{n!}\,.
\label{M2}
\end{eqnarray}
Using the relations \cite{Prudnikov1}
\begin{eqnarray}
{\rm L}_{n_0}^n(0)={n+n_0\choose n_0 }
\label{M3}
\end{eqnarray}
and
\begin{eqnarray}
\sum_{n=0}^{\infty}{\rm L}_{n_0}^n(x)\frac{z^n}{n!}=e^z{\rm L}_{n_0}(x-z)
\label{M4}
\end{eqnarray}
(${\rm L}_{n_0}^n(x)$ is the associated Laguerre polynomial), we
find that
\begin{eqnarray}
P(n_0)
=|R|^{2n_0}e^{-|R|^2|\beta|^2}{\rm L}_{n_0}(-|\beta'|^2).
\label{M5}
\end{eqnarray}
In Fig.~\ref{Fig2} the probability $P(n_0)$ is plotted for two
values of the beam-splitter transmittance $|T|$. If $|T|$ and/or
$n_{0}$ are not too small, $P(n_{0})$ as a function of $|\beta|$ can
attain a maximum, the position of which is determined by
the (positive) solution of the equation 
\begin{equation} 
\left|T\right|^2{\rm L}_{n_0}(-|\beta'|^2)=
\left|R\right|^2{\rm L}_{n_0-1}^1(-|\beta'|^2).
\end{equation}
We see that even if $|\beta|$ or $n_0$ are increased the 
probability $P(n_0)$ for detecting zero photons may increase due 
to destructive interference. In particular Fig.~\ref{Fig2}(b) reveals
that the maximum is shifted towards larger values of $|\beta|$ when $n_{0}$ is
increased. For example, assuming a highly transmitting beam splitter such 
that $|T|^{2}$ $\!=$ $\!0.99$, probabilities $P(n_0)$ of about of $10$\% 
may be possible, provided that $|\beta|$ and $n_{0}$ are sufficiently 
large ($|\beta|$ $\!\approx$ $\!40$, $n_{0}$ $\!\approx$ $\!20$).

\begin{figure}
\centering\epsfig{figure=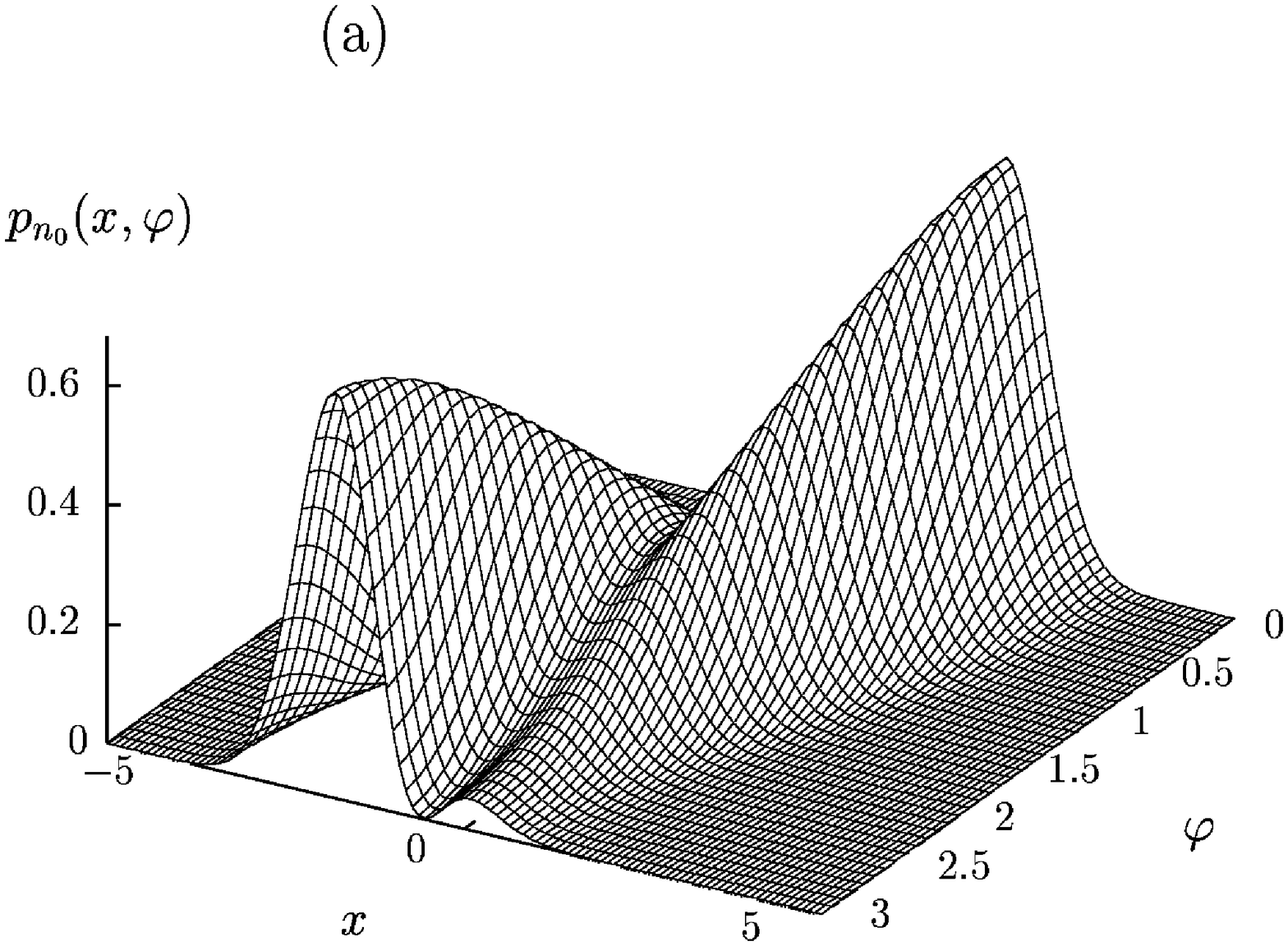,width=0.8\linewidth}

~
\vspace{0.1cm}

~

\centering\epsfig{figure=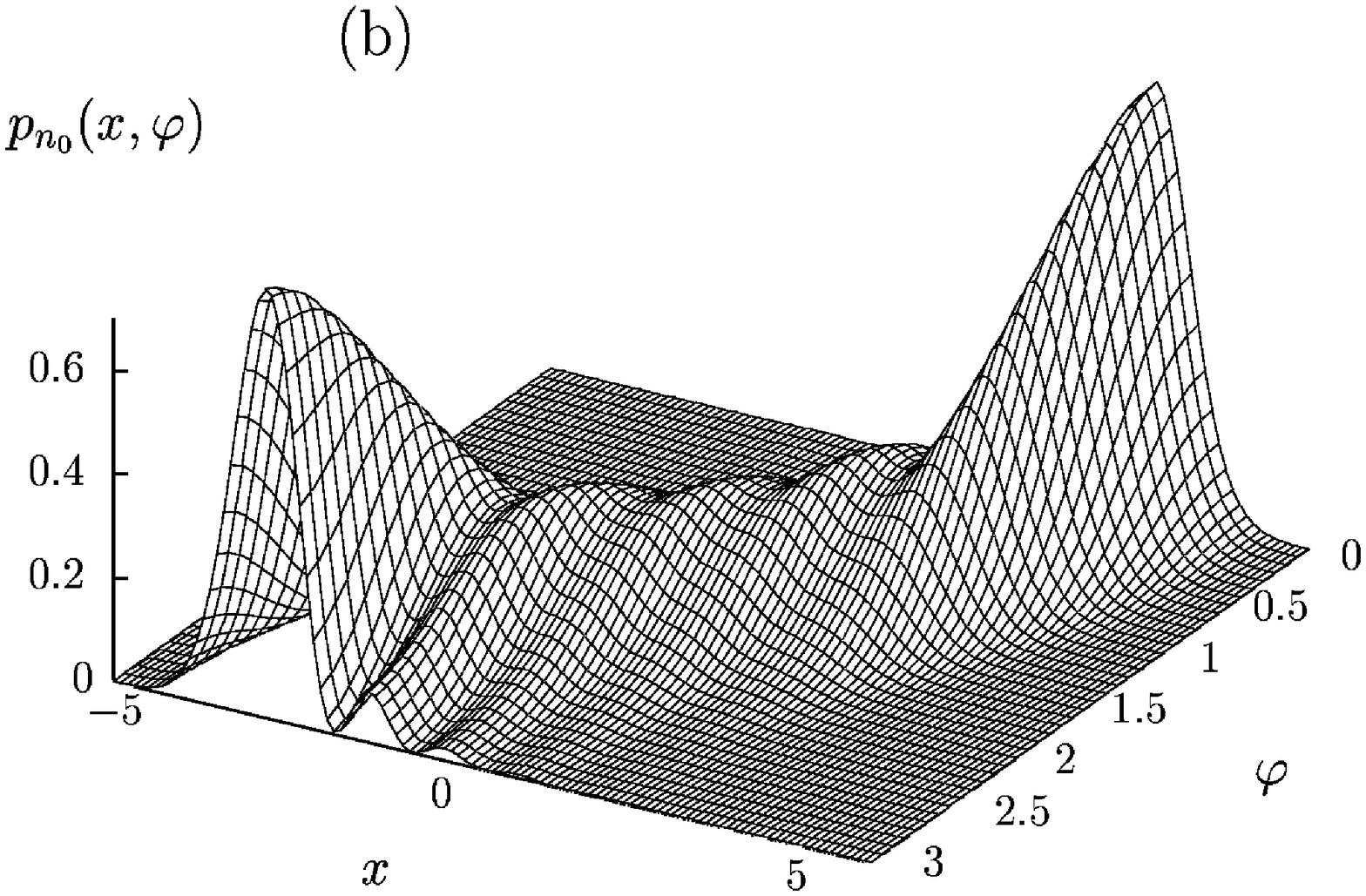,width=0.8\linewidth}
\caption{The quadrature-component distribution of photon-added
coherent states for $\beta'$ $\!=$ $\! 0.89$
($|\beta|$ $\!=$ $\! 1$, $|T|^2$ $\!=$ $\!0.8$) and two values of
$n_{0}$ [ (a) $n_{0}$ $\!=$ $\!1$; (b) $n_{0}$ $\!=$ $\!4$].
\label{Fig3}}
\end{figure}

Photon-added coherent states (\ref{Q2}) were
introduced and studied in detail in \cite{Agarwal1},
with special emphasis on their nonclassical properties,
such a squeezing and sub-Poissonian photon statistics
(for their quantum-phase statistics, see \cite{Nath1}).
In particular, in \cite{Agarwal1} analytical results for the 
Wigner and Husimi functions are given. We therefore may
restrict attention to the quadrature-component distribution
(i.e., the phase-parametrized field-strength distribution)
\begin{equation}
p_{n_0}(x,\varphi)=|\langle x,\varphi|\Psi_{n_0}\rangle|^2,
\label{F1}
\end{equation}
which can be measured in balanced homodyne detection. For this
purpose we expand the eigenvectors $|x,\varphi\rangle$ of the 
quadrature components
\begin{equation}
\hat{x}(\varphi)
= 2^{-1/2}
\left(e^{-i\varphi}\hat{a}+e^{i\varphi}\hat{a}^\dagger\right)
\label{F2}
\end{equation}
in the photon-number basis \cite{Vogel2},
\begin{equation}
|x,\varphi\rangle=(\pi)^{-1/4}
\exp\!\left(-\textstyle\frac{1}{2}x^2\right)
\sum_{n=0}^{\infty}\frac{e^{in\varphi}}
{\sqrt{2^nn!}} \, {\rm H}_n(x) |n \rangle
\label{F3}
\end{equation}
(H$_{n}$ is the Hermite polynomial). Using the identity \cite{Prudnikov1}
\begin{eqnarray}
\sum_{k=0}^{\infty}\frac{z^k}{k!}\,{\rm H}_{k+n}(x)
= \, \exp\!\left(2xz-z^2\right)
{\rm H}_{n}(x-z),
\label{F.4}
\end{eqnarray}
we find that 
\begin{eqnarray}
\lefteqn{
p_{n_0}(x,\varphi)=
\frac{2^{-n_0}}{{\cal N}'_{n_0}\sqrt{\pi}}
\exp\!\left\{
-\left[
x-\sqrt{2}|\beta'|\cos(\varphi+\varphi_{\beta'})
\right]^2
\right\}
}
\nonumber \\ && \hspace{20ex}\times \,
\left |{\rm H}_{n_0}\!\left(
x - 2^{-1/2}|\beta'|e^{i(\varphi+\varphi_{\beta'})}
\right)
\right |^2,
\label{F5}
\end{eqnarray}
where $\varphi_{\beta'}=\varphi_{\beta}+\varphi_T$. Plots of 
$p_{n_0}(x,\varphi)$ are given in Fig.~\ref{Fig3} for
$n_{0}$ $\!=$ $\!1$ [$P(n_{0})$ $\!\approx$ $\!30\%$] and
$n_{0}$ $\!=$ $\!4$ [$P(n_{0})$ $\!\approx$ $\!0.84\%$].
The $\pi$ periodic narrowing (broadening) of the quadrature-component
distribution reveals that photon-added coherent states can be regarded
as some kind of squeezed states. Nevertheless, they are in
general quite different from the two-photon coherent states \cite{Yuen2}
widely used in squeezed-light description. It is worth noting that
in contrast to the familiar two-photon coherent states the photon-added
coherent states are non-Gaussian states. In particular, the variance of
$\hat{x}(\varphi)$ is given by \cite{Agarwal1}
\begin{eqnarray}
\lefteqn{
(\Delta\hat{x}(\varphi))^2=\langle\hat{x}(\varphi)^2\rangle-
\langle\hat{x}(\varphi)\rangle^2
}
\nonumber \\ && \hspace{1ex}= \,
\frac{1}{2{\rm L}_{n_0}(-2|\beta'|^2)^2}\bigg\{
2|\beta'|^2\big[{\rm L}_{n_0}^2(-|\beta'|^2){\rm L}_{n_0}(-|\beta'|^2)
\nonumber \\ && \hspace{1ex}-\,
{\rm L}_{n_0}^1(-|\beta'|^2)^2\big]\cos(2(\varphi+\varphi_{\beta'}))
-2|\beta'|^2{\rm L}_{n_0}^1(-|\beta'|^2)^2
\nonumber \\ && \hspace{1ex}-\,
{\rm L}_{n_0}(-|\beta'|^2)^2+2(n_0+1){\rm L}_{n_0}(-|\beta'|^2)
{\rm L}_{n_0+1}(-|\beta'|^2)\bigg\}.
\end{eqnarray}
With increasing number of added photons, $n_{0}$, the squeezing effect
is (for chosen $|\beta|'$ $\!>$ $\!0$) enhanced. Needless to say
that when $n_{0}$ $\!=$ $\!0$ (i.e., when no photons are added), then the
conditional state $|\Psi_{0}\rangle$ simply reduces to the coherent
state $|\beta'\rangle$. Finally, it should be noted that when the vacuum
is mixed with a Fock state and the zero-photon measurement in one of the
output channels of the beam splitter is replaced with a measurement of
the $Q$ function (in perfect six- or eight-port balanced homodynings), then
the conditional states of the other output channel are similar to
the photon-added coherent states \cite{Ban3}.


\subsection{Squeezed vacuum states}
\label{sec3b}

Let us now consider a squeezed-vacuum input state
\begin{equation}
|\Phi\rangle = |0\rangle_{\rm s} = \hat{S}(\xi)| 0 \rangle,
\label{1.61a}
\end{equation}
where
\begin{eqnarray}
\hat{S}(\xi)| 0 \rangle =
\exp\!\left\{-\textstyle\frac{1}{2}
\left[(\xi\hat{a}^{\dagger})^{2} - \xi^{\ast}\hat{a}^{2}\right]
\right\} | 0 \rangle
= (1-|\kappa|^{2})^{1/4}\sum_{n=0}^{\infty}
\frac{[(2n)!]^{1/2}}{2^{n}\,n!}\,
\kappa^{n}|2n\rangle,
\label{1.2}
\end{eqnarray}
with $\xi$ $\!=$ $\!|\xi|e^{i\varphi_{\xi}}$ and
$\kappa$ $\!=$ $\!e^{i\varphi_{\xi}}\tanh|\xi|$. The conditional output 
states (\ref{1.4m}) are seen to be photon-added squeezed vacuum states
\begin{eqnarray}
|\Psi_{n_0}\rangle 
=\frac{1}{\sqrt{{\cal N}'_{n_0}}}
(\hat{a}^\dagger)^{n_0}\hat{S}(\xi')| 0 \rangle ,
\label{1.4a}
\end{eqnarray}
where $\xi'$ $\!=$ $\!|\xi'|e^{i(\varphi_{\xi}+2\varphi_T)}$,
and $\tanh|\xi'|$ $\!=$ $\!|T|^2\tanh|\xi|$. In the photon-number
basis they read as
\begin{eqnarray}
|\Psi_{n_0}\rangle
=\frac{(1-|\kappa'|^{2})^{1/4}}{\sqrt{{\cal N}'_{n_0}}}
\sum_{n=n_0}^{\infty} b_{n,n_0}(\kappa') \, |n\rangle,
\label{M7}
\end{eqnarray}
where
\begin{eqnarray}
b_{n,n_0}(\kappa')=\frac{\sqrt{n!}}
{\Gamma\!\left[\frac{1}{2}(n-n_0)+1\right]}
\,{\textstyle\frac{1}{2}\left[1+(-1)^{n-n_0}\right]}
\left(\textstyle\frac{1}{2}\kappa'\right)^{(n-n_0)/2}
\label{M8}
\end{eqnarray}
and $\kappa'$ $\!=$ $\!T^2\kappa$. Note that when $n_{0}$ $\!=$ $\!0$
then $|\Psi_{n_{0}}\rangle$ simply reduces to a squeezed vacuum state
(\ref{1.2}), with $\kappa'$ in place of $\kappa$.

\begin{figure}
\centering\epsfig{figure=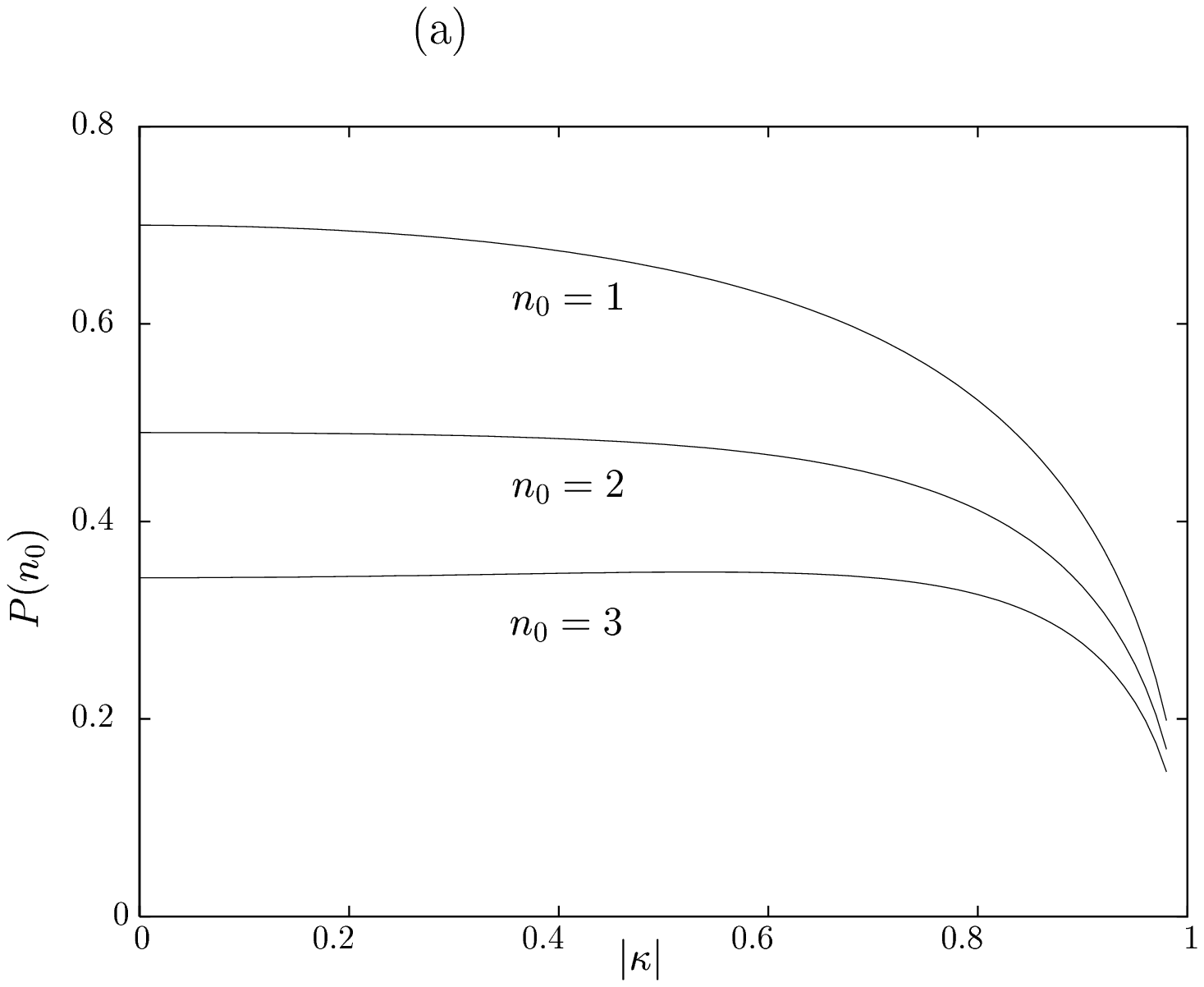,width=0.8\linewidth}

~

\vspace{0.2cm}

~

\centering\epsfig{figure=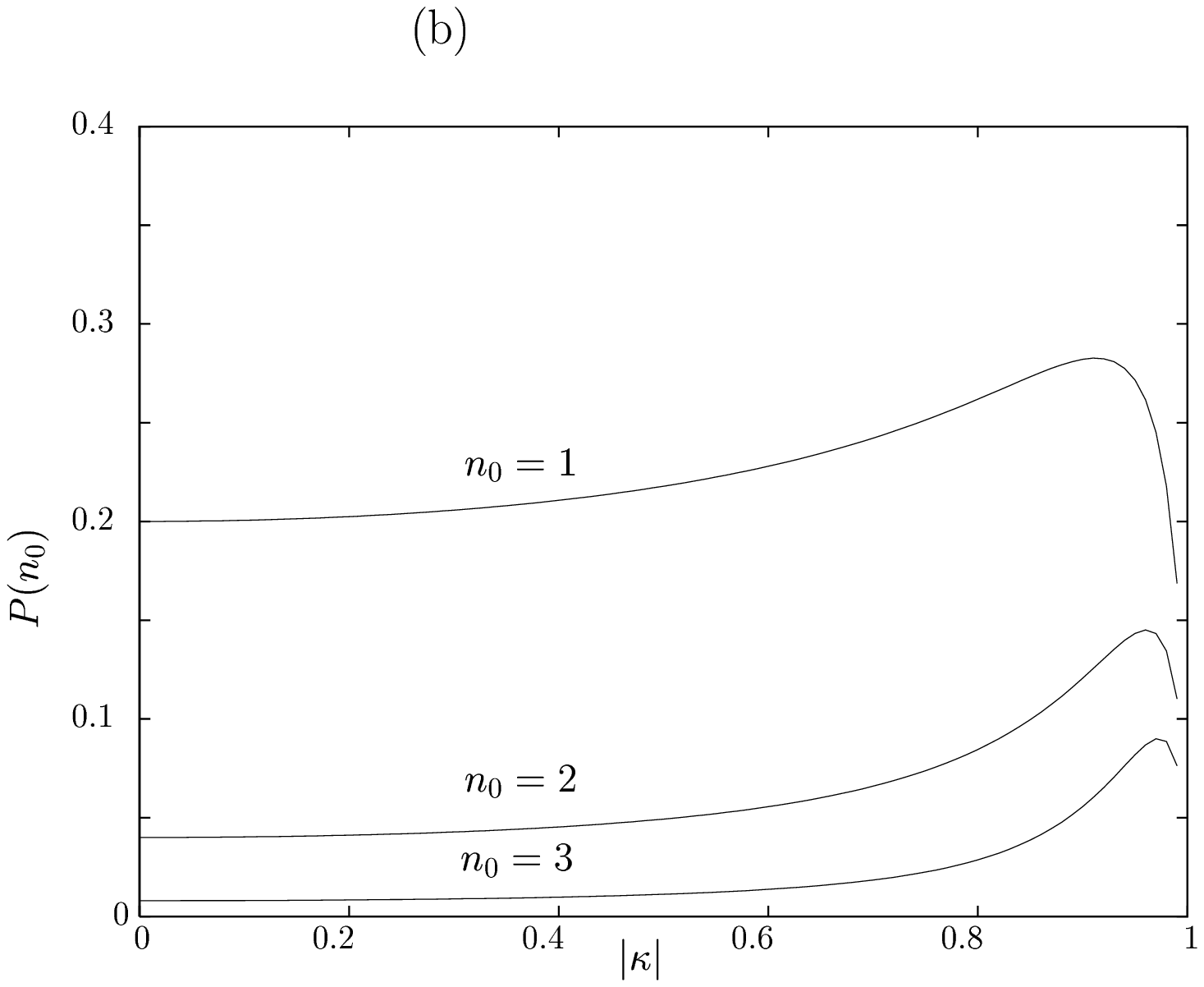,width=0.8\linewidth}
\caption{The probability of producing photon-added squeezed vacuum
states is shown as a function of $|\kappa|$ for two values of the
beam-splitter transmittance [(a) $|T|^2$ $\!=$ $\!0.3$;
(b) $|T|^2$ $\!=$ $\!0.8$] and various values of $n_{0}$.
\label{Fig4}}
\end{figure}

For the sake of tranceparency and without loss of generality we will
assume that $\kappa'$ is real,
i.e., $\!\varphi_{\xi}+2\varphi_T$ $\!=$ $\!k\pi$, with $k$ being an
integer. Note that the effect of other phases is simply a rotation in
phase space. Using the doubling formula for the Gamma-function,
\begin{equation} 
\frac{\Gamma(2n)}{\Gamma(n+1/2)}
= \frac{4^n}{2\sqrt{\pi}}\,\Gamma(n),
\label{B6}
\end{equation}
and the Gauss-series of the hypergeometric function \cite{Erdelyi1},
\begin{eqnarray}
{\rm F}(a,b,c;z)=\frac{\Gamma(c)}{\Gamma(a)\Gamma(b)}
\sum_{n=0}^{\infty}\frac{\Gamma(a+n)\Gamma(b+n)}{\Gamma(c+n)}\frac{z^n}{n!},
\label{B7}
\end{eqnarray}
the normalization constant ${\cal N}'_{n_0}$ can be derived to be
\begin{eqnarray}
{\cal N}'_{n_0}
= \sqrt{1\!-\!\kappa'^2}\,n_0!\,
{\rm F}\!\left[\textstyle\frac{1}{2}(n_0+1),
\textstyle\frac{1}{2}(n_0+2),1;\kappa'^2\right].
\label{1.3b}
\end{eqnarray}
   From Eqs.~(\ref{M1}) and (\ref{1.2}) [together with
Eqs.~(\ref{B6}) and (\ref{B7})] the probability of producing
photon-added squeezed vacuum states is derived to be
\begin{eqnarray}
P(n_0) = |R|^{2n_0} \sqrt{1\!-\!|\kappa|^2}\,
{\rm F}\!\left(n_0+1,\textstyle\frac{1}{2},1;\kappa'^2\right).
\label{M13a}
\end{eqnarray}
Examples of $P(n_0)$ are plotted in Fig.~\ref{Fig4}. In
particular we see that for not too small transmittance of the beam
splitter (and chosen $n_{0}$) the probability $P(n_0)$ can increase
with the value of $|\kappa|$, which is similar to the behavior shown
in Fig.~\ref{Fig3} for a coherent input state. Note that the mean
photon number of the squeezed vacuum (\ref{1.2}) is given by
$|\kappa|^2/(1-|\kappa|^2)$.

\begin{figure}
\centering\epsfig{figure=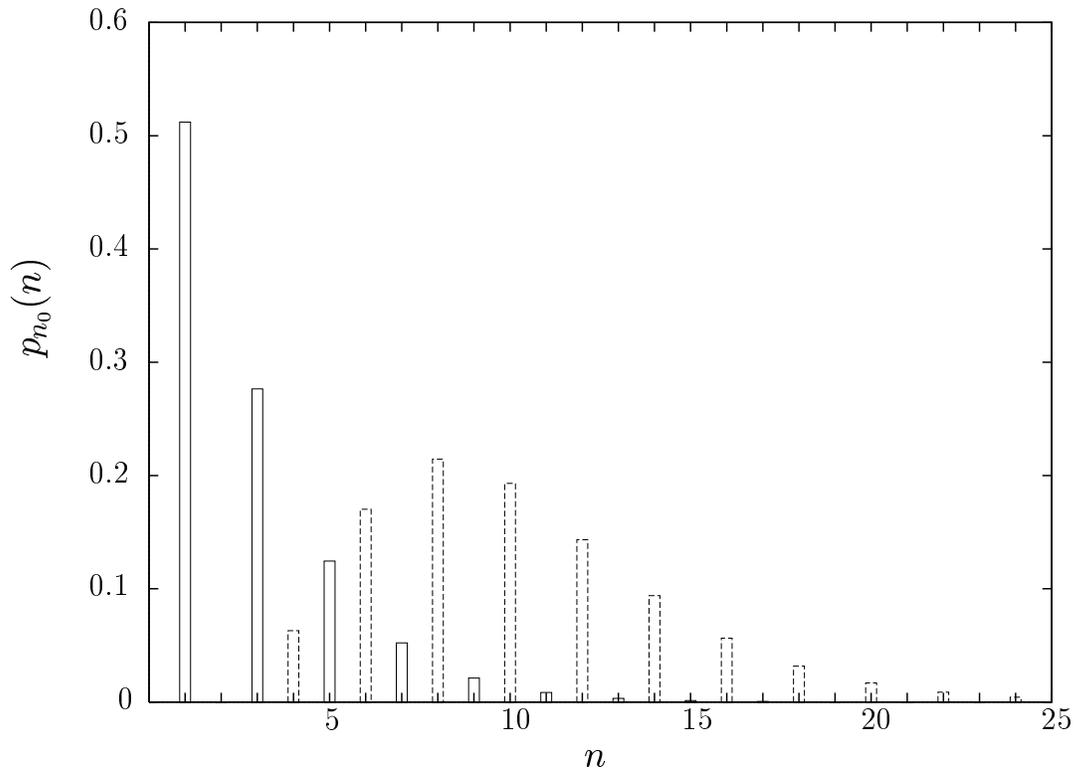,width=1\linewidth}
\caption{The photon-number distribution of photon-added
squeezed vacuum states for $\kappa'$ $\!=$ $\! 0.6$
($|\kappa|$ $\!=$ $\! 0.67$, $|T|^2$ $\!=$ $\!0.8$) and two values of
$n_{0}$ ($n_{0}$ $\!=$ $\!1$, full bars; $n_{0}$ $\!=$ $\!4$, dashed bars).
\label{Fig5}}
\end{figure}

Using Eq.~(\ref{M7}), the photon-number distribution
\begin{equation}
p_{n_0}(n) = |\langle\,n|\Psi_{n_0}\rangle|^2
\label{n1}
\end{equation}
of the photon-added squeezed vacuum states can be given by
\begin{equation}
p_{n_0}(n) = \frac{1}{{\cal N}''_{n_0}}
\left| b_{n,n_0}(\kappa') \right|^{2} \quad {\rm if} \quad n\geq n_0
\label{n2}
\end{equation}
and $p_{n_0}(n)$ $\!=$ $\!0$ elsewhere, and
\begin{eqnarray}
{\cal N}''_{n_0}
= n_0!\,{\rm F}\!\left[\textstyle\frac{1}{2}(n_0+1),
\textstyle\frac{1}{2}(n_0+2),1;\kappa'^2\right].
\label{1.3bbb}
\end{eqnarray}
   From Eqs.~(\ref{n2}) and (\ref{M8}) we easily see that when the
number of the added photons, $n_0$, is even (odd), then the
photon-number distribution is nonzero only for even (odd) photon
numbers. This oscillating behavior of the photon-number
distribution obviously reflects the fact that only even photon numbers
contribute to the squeezed vacuum state to which photons are added.
In particular the mean photon number
\begin{equation}
\langle\hat{n}\rangle = \sum_{n} n\,p_{n_0}(n)
\label{n3}
\end{equation}
may be rewritten as
\begin{eqnarray}
\langle\hat{n}\rangle=
\kappa'\frac{\partial}{\partial\kappa'}
\log\!\left[\kappa'^{n_0}{\cal N}''_{n_0}\right] .
\label{M9}
\end{eqnarray}
Using standard formulas for the derivative of the hypergeometric
function \cite{Erdelyi1}, we find that
\begin{eqnarray}
\langle\hat{n}\rangle=
n_0+{\textstyle\frac{1}{2}}\kappa'^2(n_0+1)(n_0+2)
\,\frac{{\rm F}\!\left[ \frac{1}{2}(n_0+3),
\frac{1}{2}(n_0+4),2;\kappa'^2\right]}
{{\rm F}\!\left[\frac{1}{2}(n_0+1),
\frac{1}{2}(n_0+2),1;\kappa'^2\right]} \, .
\label{M10}
\end{eqnarray}
As expected, for $\kappa'$
$\!\to$ $\!0$ $\langle\hat{n}\rangle$ approaches $n_0$ and it
increases with $\kappa'$ and $n_0$. Examples of the photon-number
distribution are shown in Fig.~\ref{Fig5} for
$n_{0}$ $\!=$ $\!1$ [$P(n_{0})$ $\!\approx$ $\!23\%$] and
$n_{0}$ $\!=$ $\!4$ [$P(n_{0})$ $\!\approx$ $\!0.45\%$].

\begin{figure}
\centering\epsfig{figure=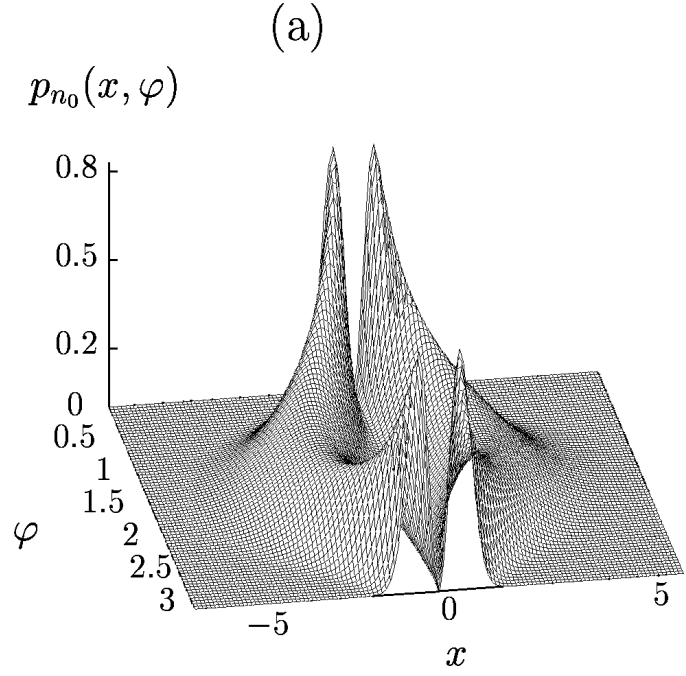,width=0.65\linewidth}

~

\vspace{0.05cm}

~

\centering\epsfig{figure=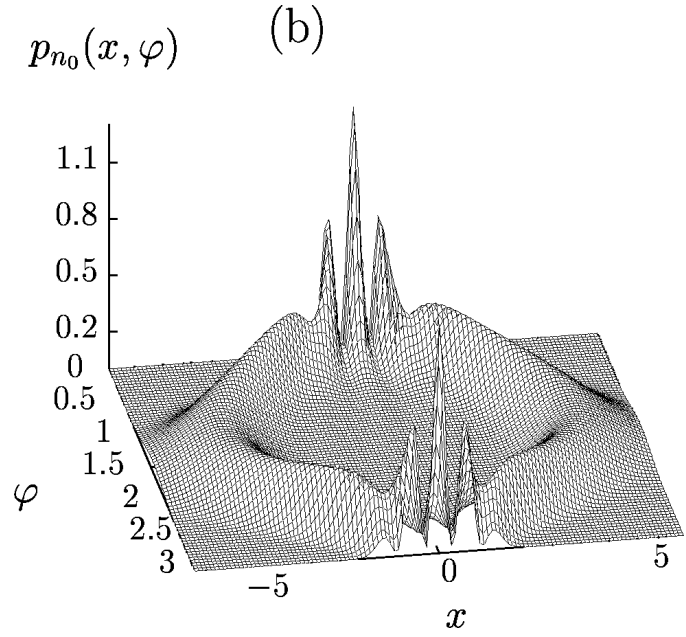,width=0.65\linewidth}
\caption{The quadrature-component distribution of photon-added
squeezed vacuum states for $\kappa'$ $\!=$ $\! 0.6$
($|\kappa|$ $\!=$ $\! 0.67$, $|T|^2$ $\!=$ $\!0.8$) and two values of
$n_{0}$ [(a) $n_{0}$ $\!=$ $\!1$; (b) $n_{0}$ $\!=$ $\!4$].
\label{Fig6}}
\end{figure}

Let us now turn to the quadrature-component distribution.
Combining Eqs.~(\ref{F1}), (\ref{F3}), and (\ref{M7}),
we derive
\begin{eqnarray}
p_{n_0}(x,\varphi)=
\frac{2^{-n_0}}{{\cal N}''_{n_0} \sqrt{\pi \Delta^{n_0+1}}}
\,\exp\!\left( - \frac{1-\kappa'^{2}}{\Delta} \, x^{2} \right)
\left |
{\rm H}_{n_0}\!\left[
\sqrt{(1-\kappa' e^{i2\varphi})/\Delta}\,x
\right]\right |^2 ,
\label{F4}
\end{eqnarray}
where
\begin{equation}
\Delta = 1 + \kappa'^2 - 2 \kappa' \cos(2\varphi).
\label{M12}
\end{equation}
Note that in the derivation of Eq.~(\ref{F4}) the summation formula
\cite{Prudnikov1}
\begin{eqnarray}
\lefteqn{
\sum_{k=0}^{\infty}\frac{z^k}{k!}\,{\rm H}_{2k+n}(x)=
(1+4z)^{-n/2-1/2}
}
\nonumber \\ && \hspace{2ex} \times \,
\exp\!\left(\frac{4zx^2}{1+4z}\right)
{\rm H}_{n}\!\left(\frac{x}{\sqrt{1+4z}}\right)
\quad \left(|z| < \textstyle\frac{1}{2}\right)
\label{M11}
\end{eqnarray}
has been used.
The quadrature-component distributions plotted in Fig.~\ref{Fig6}
correspond to the same parameters as in Fig.~\ref{Fig5}. From inspection
of Fig.~\ref{Fig6}, interference fringes for $\varphi$ close to $0$ or
$\pi$ are seen, whereas for $\varphi$ near $\pi/2$ two separated peaks
are observed. The behavior is typical of a Schr\"{o}dinger-cat-like
superposition of two macroscopically distinguishable states.

\begin{figure}
\centering\epsfig{figure=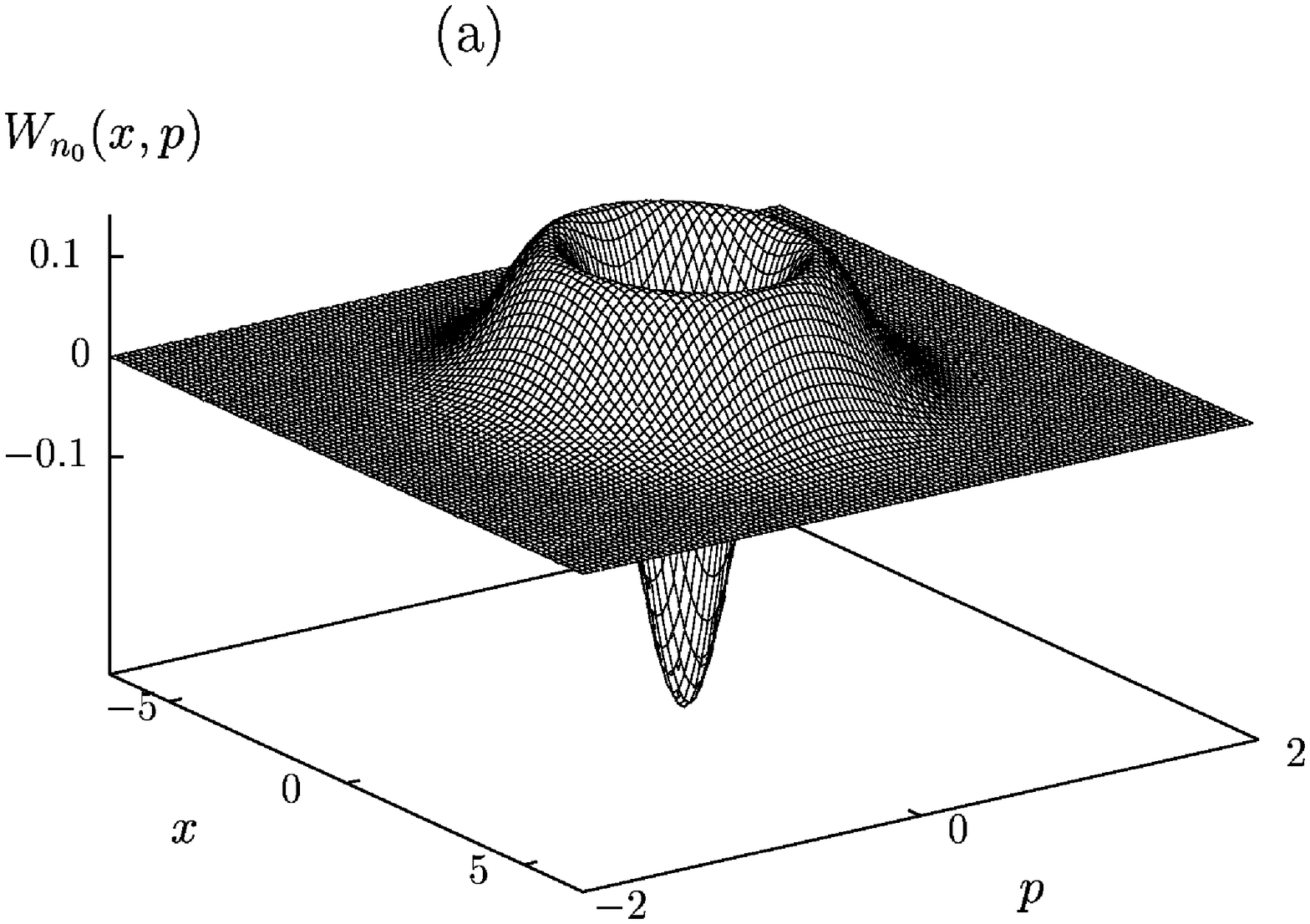,width=0.8\linewidth}

~
\vspace{0.1cm}

~

\centering\epsfig{figure=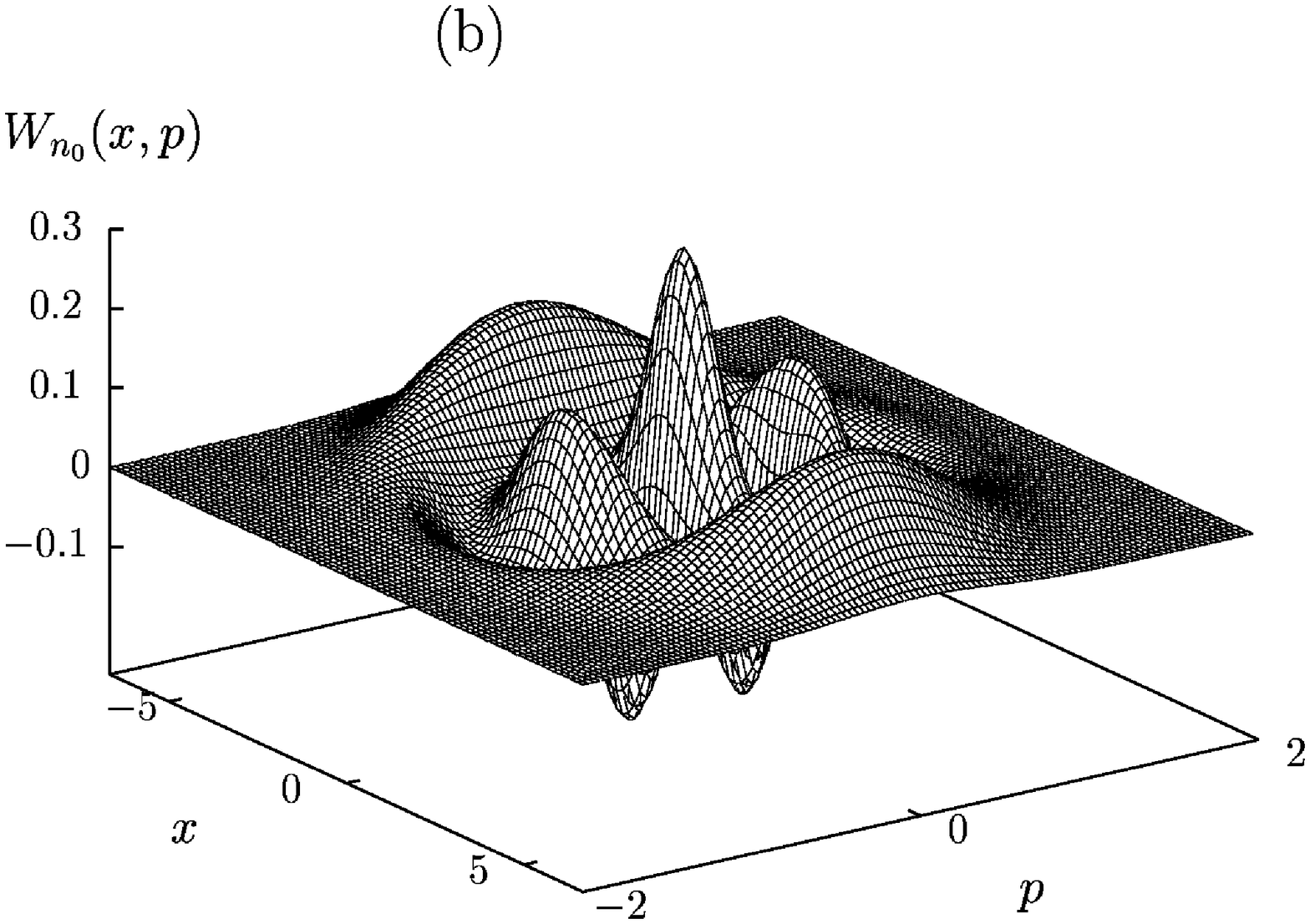,width=0.8\linewidth}
\caption{The Wigner function of photon-added
squeezed vacuum states for $\kappa'$ $\!=$ $\! 0.6$
($|\kappa|$ $\!=$ $\! 0.67$, $|T|^2$ $\!=$ $\!0.8$) and two values of
$n_{0}$ [(a) $n_{0}$ $\!=$ $\!1$; (b) $n_{0}$ $\!=$ $\!4$].
\label{Fig7}}
\end{figure}

Next let us calculate the Wigner function
\begin{eqnarray}
\label{W1}
W_{n_0}(x,p) = \frac{1}{\pi}\int\limits_{-\infty}^{+\infty} {d}y \, e^{2ipy}
\langle x\!-\!y|\Psi_{n_0}\rangle\langle\Psi_{n_0}|x\!+\!y\rangle.
\end{eqnarray}
Using Eqs.~(\ref{F3}) and (\ref{M7}) [together with Eq.~(\ref{M11})],
after some calculation we obtain
\begin{eqnarray}
\lefteqn{
W_{n_0}(x,p)= 
\frac{ 2e^{-\lambda x^2}}{\pi^{3/2}
{\cal N}''_{n_0} [2(\kappa'+1)]^{n_0+1}}
}
\nonumber \\ && \hspace{8ex} \times
\int\limits_{-\infty}^{+\infty} {d}y
e^{-\lambda y^2 + 2i p y}\,
{\rm H}_{n_0}\!\left(\frac{x\!-\!y}{\sqrt{1\!+\!\kappa'}}\right)
{\rm H}_{n_0}\!\left(\frac{x\!+\!y}{\sqrt{1\!+\!\kappa'}}\right),
\label{W2}
\end{eqnarray}
where
\begin{equation}
\lambda = \frac{1-\kappa'}{1+\kappa'} \,.
\label{W3}
\end{equation}
Performing the $y$ integration \cite{Prudnikov1} yields
\begin{eqnarray}
\label{W4}
\lefteqn{
W_{n_0}(x,p)=
\!\frac{|\kappa'|^{n_0}\sqrt{2}}
{\pi{\cal N}''_{n_0}[2(1\!-\!\kappa'^2)]^{n_0+1/2}}
\exp\!\left(\! - \lambda x^2 \! - \! \frac{p^2}{\lambda} \!\right)
}
\nonumber \\ && \hspace{8ex} \times
\sum_{k=0}^{n_0}\!{n_0\choose k}^2 \!k!
\!\left(\frac{-2}{|\kappa'|}\right)^{k}
\left |{\rm H}_{n_0-k}\!\left[i\sqrt{\frac{\lambda}{\kappa'}}
\left(x\!+\!i\frac{p}{\lambda}\right)\right]\right |^2\!\!.
\end{eqnarray}
The Wigner functions in Fig.~\ref{Fig7} are plotted for the same
parameters as in Figs.~\ref{Fig5} and \ref{Fig6}. We again
recognize the typical features of Schr\"odinger-cat-like states.

Finally, let us consider the Husimi function
\begin{equation}
Q_{n_0}(x,p)= \frac{1}{2\pi} \, |\langle\alpha|\Psi_{n_0}\rangle|^2 ,
\label{M13}
\end{equation}
where $|\alpha \rangle$ is a coherent state and
$\alpha$ $\!=$ $\!2^{-1/2}(x\!+\!ip)$. Recalling the expansion
of the coherent states in the Fock basis, Eq.~(\ref{Q1}), and
using Eq.~(\ref{M7}), we easily find that
\begin{eqnarray}
\label{Q3}
Q_{n_0}(x,p)=\frac{(x^2+y^2)^{n_0}}{\pi 2^{n_0+1}{\cal N}''_{n_0}}
\,\exp\!\left\{ -\textstyle\frac{1}{2}
\left[ (1-\kappa')x^2+(1+\kappa')p^2 \right] \right\}.
\end{eqnarray}
Note that the Husimi function is a phase-space function that can be
measured in multiport balanced homodyning, such as six-port \cite{Walker1}
or eight-port \cite{Walker2} detections. Since the Husimi function can be
regarded as a smoothed Wigner function, the oscillating behavior
and the negative values that are typical of the Wigner
function (see Fig.~\ref{Fig7}) cannot be observed.

Schr\"{o}dinger cat-like states are commonly defined as superpositions of
two macroscopically distinguishable states. From Eqs.~(\ref{M7}) and
(\ref{M8}) it is seen that $|\Psi_{n_0}\rangle$ can be given by
the superposition
\begin{eqnarray}
\label{Co1}
|\Psi_{n_0}\rangle = A
\left(|\Psi^{(+)}_{n_0}\rangle +|\Psi ^{(-)}_{n_0}\rangle\right) ,
\end{eqnarray}
where 
\begin{equation}
|\Psi ^{(\pm)}_{n_0}\rangle
=\frac{1}{\sqrt{{{\cal N}''^{(\pm)}_{n_0}}}}
\sum_{n=n_0}^{\infty}
b^{(\pm)}_{n,n_0}(\kappa') \, |n\rangle ,
\label{Co2}
\end{equation}
with
\begin{equation}
b^{(\pm)}_{n,n_0}(\kappa')=\frac{\sqrt{n!}}
{\Gamma[ (n-n_0)/2 + 1 ]}
\left(\pm\sqrt{\textstyle\frac{1}{2}\kappa'}\right)^{n-n_0}\!\!\!.
\label{Co3}
\end{equation}
Using standard relations \cite{Erdelyi1}, the normalization factor
${{\cal N}''^{(\pm)}_{n_0}}$ in Eq.~(\ref{Co2}) can be expressed in
terms of derivatives of the hypergeometric function,
\begin{eqnarray}
{{\cal N}''^{(\pm)}_{n_0}} = \frac{\partial^{n_0}}
{\partial{\kappa'}^{n_0}}{\kappa'}^{n_0}\bigg\{
{\rm F}\!\left(\textstyle\frac{1}{2},1,1;\kappa'^2\right)
+ \frac{2}{\pi}\kappa'
{\rm F}\!\left(1,1,\textstyle\frac{3}{2};\kappa'^2\right)\bigg\}\,,
\label{M15}
\end{eqnarray}
and the normalization constant in Eq.~(\ref{Co1}) is given by
$A$ $\!=$ $\!\frac{1}{2}({{\cal N}''^{(\pm)}_{n_0}}/{\cal N}''_{n_0})^{1/2}$.

\begin{figure}
\centering\epsfig{figure=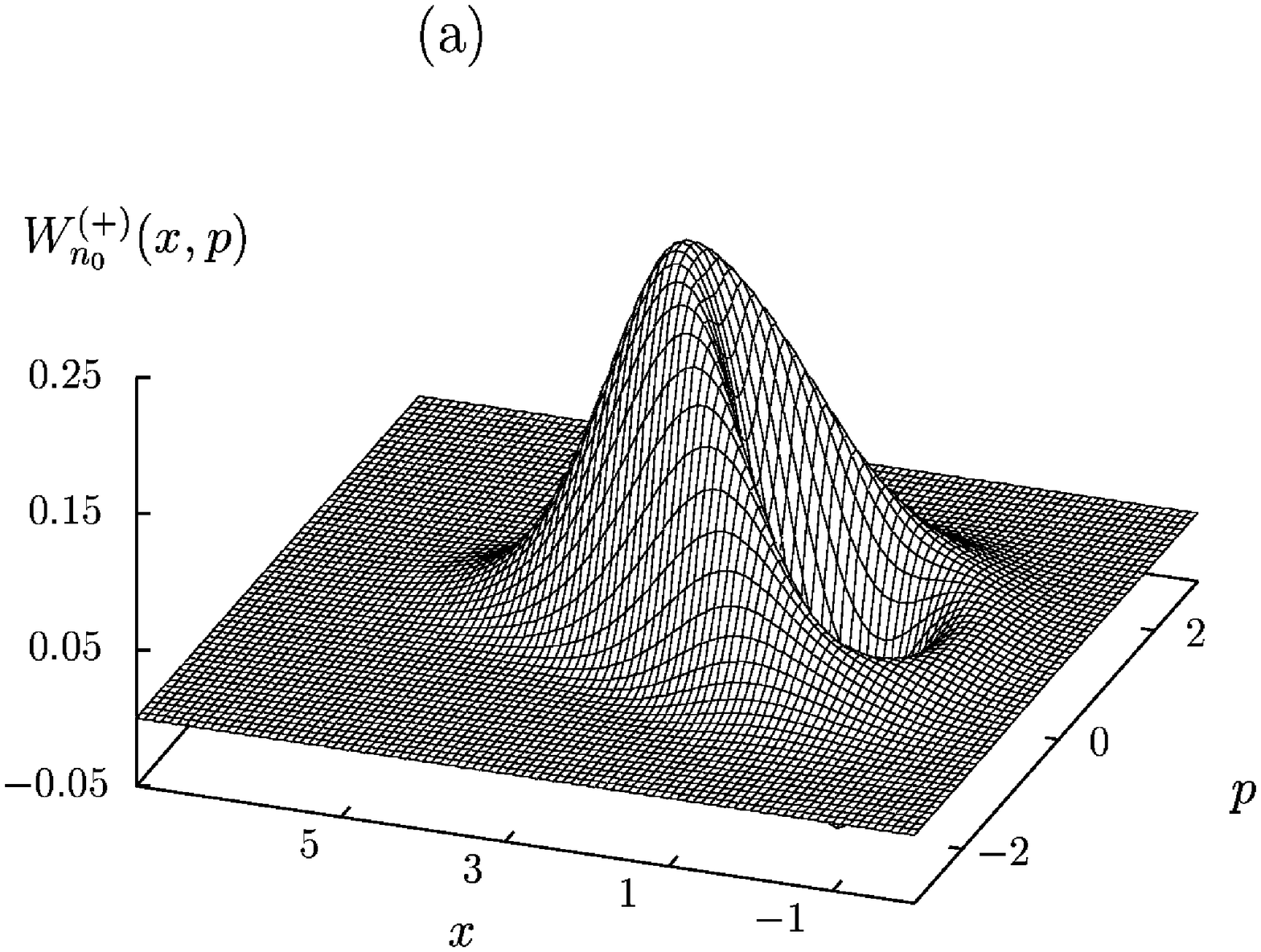,width=0.6\linewidth}

~
\vspace{0.1cm}
~

\centering\epsfig{figure=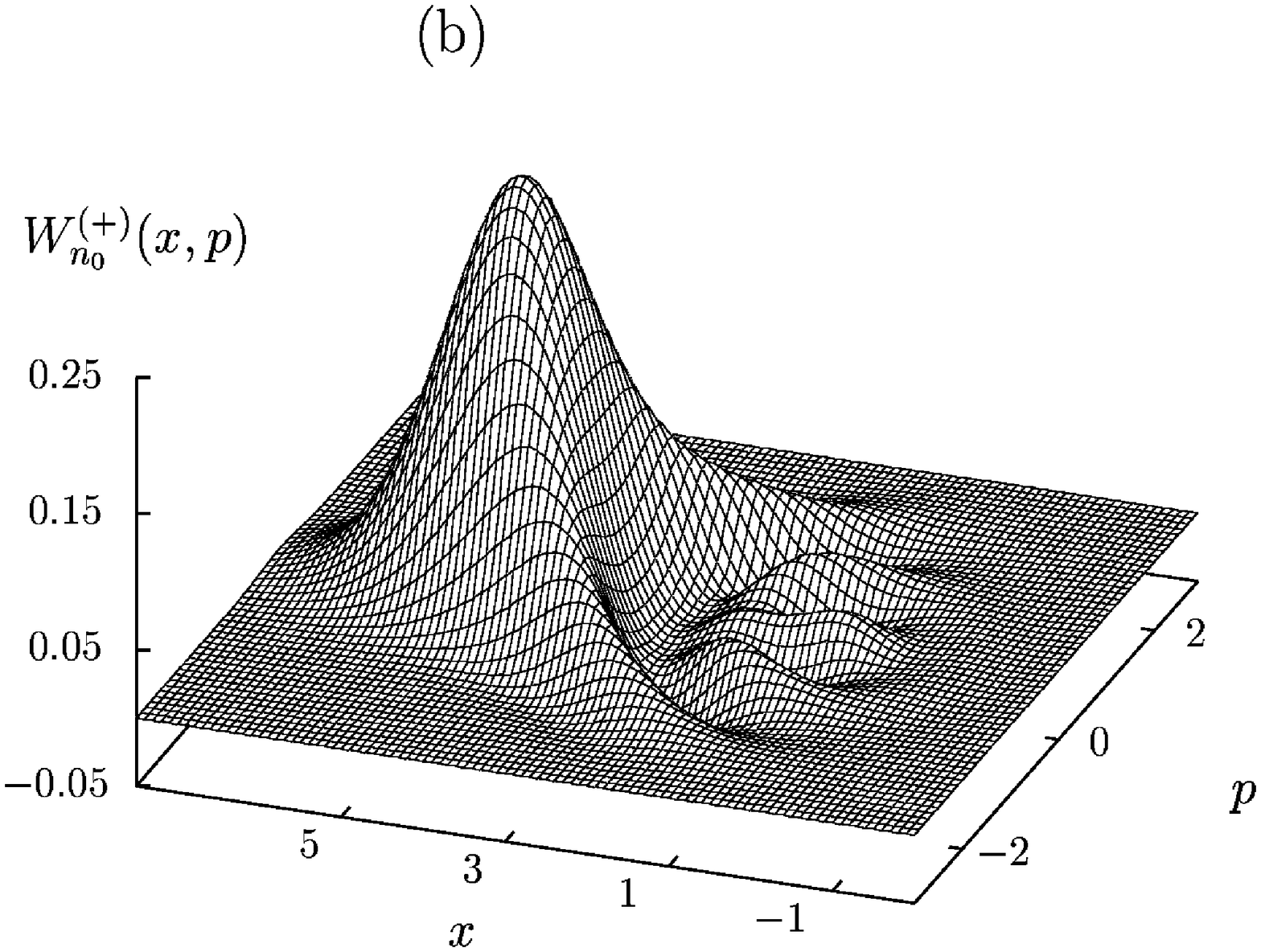,width=0.6\linewidth}

~
\vspace{0.1cm}

~
\centering\epsfig{figure=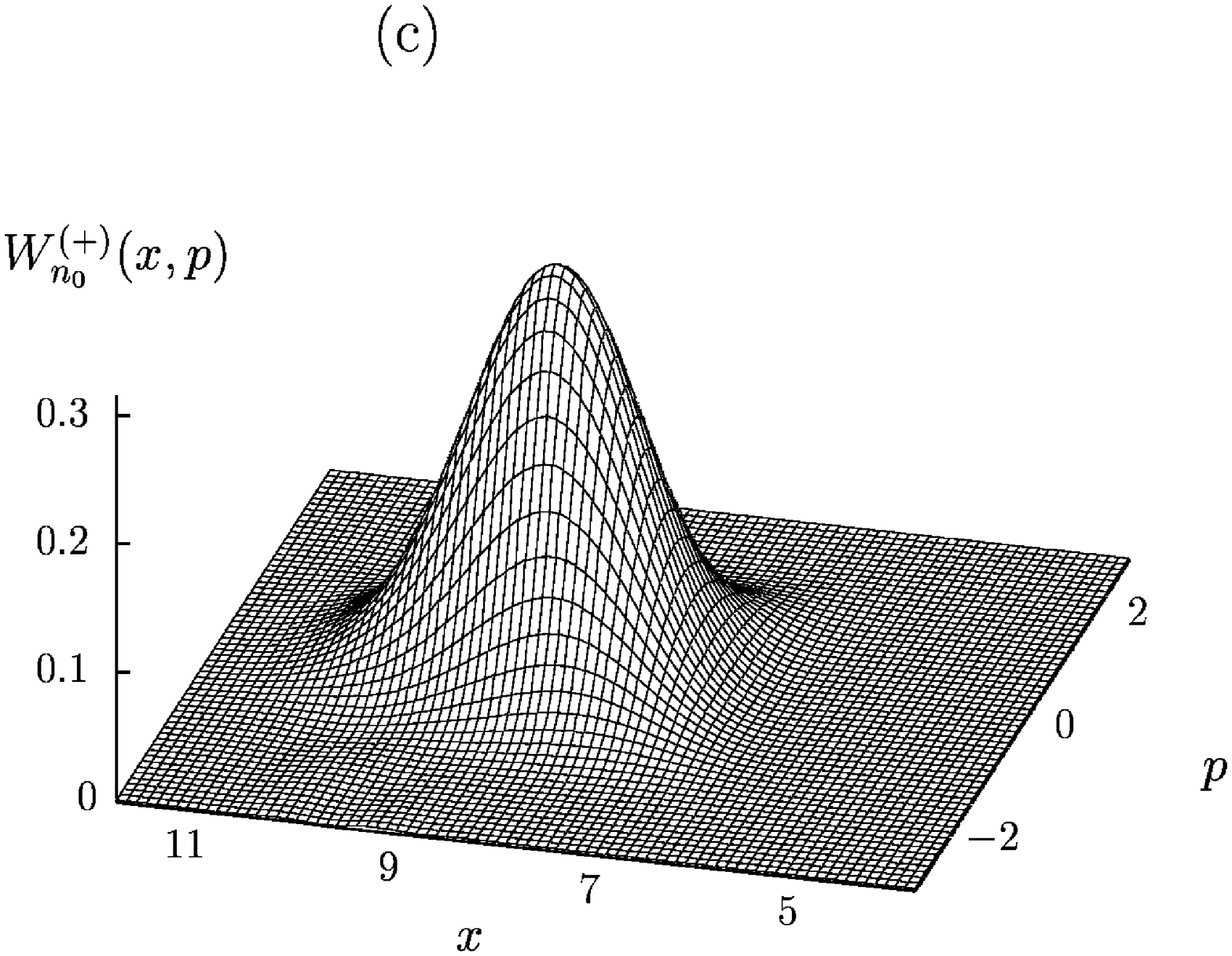,width=0.6\linewidth}
\caption{The Wigner function of the component states
$|\Psi_{n_{0}}^{(+)}\rangle$ for $\kappa'$ $\!=$ $\! 0.6$
($|\kappa|$ $\!=$ $\! 0.67$, $|T|^2$ $\!=$ $\!0.8$) and three values of
$n_{0}$ [(a) $n_{0}$ $\!=$ $\!1$; (b) $n_{0}$ $\!=$ $\!4$;
(c) $n_{0}$ $\!=$ $\!15$].
\label{Fig8}}
\end{figure}

To demonstrate the nonclassical properties of the component states
$|\Psi^{(\pm)}_{n_0}\rangle$, in Fig.~\ref{Fig8} we have plotted
the Wigner function $W_{n0}^{(+)}(x,p)$ of $|\Psi^{(+)}_{n_0}\rangle$
for various values of $n_{0}$. We see that with increasing $n_{0}$
the Wigner function becomes less structurized and the negative
values are more and more suppressed [e.g., for $\kappa'$ $\!=$ $\!0.6$
and $n_0$ $\!=$ $\! 15$ the Wigner function attains negative values of
the order of magnitude of $-10^{-4}$, Fig.~\ref{Fig8}(c)].
It is worth noting that the component states $|\Psi ^{(\pm)}_{n_0}\rangle$
can be regarded as non-Gaussian squeezed coherent states that
tend to the familiar Gaussian squeezed coherent states as $n_{0}$
becomes sufficiently large. To illustrate the difference between
the states $|\Psi^{(\pm)}_{n_0}\rangle$ and the Gaussian squeezed
coherent states, let us consider the  Husimi function $Q_{n_0}^{(\pm)}(x,p)$ 
$\!$ $\!=$ $\!|\langle\alpha|\Psi_{n_0}^{(\pm)}\rangle|^2$,
with $\alpha$ $\!=$ $\!2^{-1/2}(x\!+\!ip)$. Using the expansions
(\ref{Q1}) and (\ref{Co2}), we derive
\begin{eqnarray}
Q_{n_0}^{(\pm)}(x,p) =
\frac{|\alpha|^{2n_0}e^{-|\alpha|^2}}{2\pi{{\cal N}''_{n_0}}^{(\pm)}}
\,\exp\!\left[{\textstyle\frac{1}{2}}
\kappa'\left(\alpha^2+{\alpha^\ast}^2\right)\right]
\left |{\rm Erfc}\!\left(\mp
\sqrt{\textstyle\frac{1}{2}\kappa'}\alpha\right)\right|^2.
\label{M20}
\end{eqnarray}
Here, the relation \cite{Abramowitz1}
\begin{eqnarray}
\sum_{n=0}^{\infty}\frac{z^n}{\Gamma(\frac{1}{2}n+1)}=
\exp(z^2)\,{\rm Erfc}(-z)
\end{eqnarray}
has been used, where ${\rm Erfc}(z)$ is the complementary
complex error function defined by
\begin{eqnarray}
{\rm Erfc}(z)=1-{\rm Erf}(z)
=\frac{2}{\sqrt{\pi}}\int_{z}^{\infty} {d}z \, e^{-t^2} .
\end{eqnarray}
   From Eq.~(\ref{M20}) the asymptotic form of $Q_{n_0}^{(\pm)}(x,p)$
for large $n_{0}$ is derived to be (see \ref{app1})
\begin{eqnarray}
Q_{n_0}^{(\pm)}(x,p)\approx\frac{n_0!}{4\pi^2\!n_0{{\cal N}''_{n_0}}^{(\pm)}}
\,\exp(-|\alpha\mp\sqrt{n_0}|^2)
\exp\!\left[{\textstyle\frac{1}{2}\kappa'}
\left(\alpha^2+{\alpha^\ast}^2\right)\right].
\label{M21}
\end{eqnarray}
   From inspection of  Eq.~(\ref{M21}) we see that for large $n_0$
the Husimi function $Q_{n_0}^{(\pm)}(x,p)$ becomes a single-peaked
Gaussian centered at $\pm\sqrt{n_0}$, i.e., when 
$n_{0}$ $\!\to$ $\!\infty$ then the states $|\Psi_{n_0}^{(\pm)}\rangle$ 
tend to the familiar Gaussian squeezed coherent states.
 

\section{Mixed photon-added states}
\label{sec4}

So far we have assumed that a mode prepared in a photon-number state 
$|n_{0}\rangle$ is fed into one of the input ports of a beam splitter, so 
that exactly $n_{0}$ photons can be added to the state of the (signal) mode
fed into the other input port. In practice however, it may be more 
realistic to consider a statistical mixture of photon-number states 
rather than a pure state, because of smoothings in the generation
of photon-number states \cite{Davidovich1}. It is worth noting
that -- apart from some smearing -- the above given results remain valid 
as long as the statistical mixture of photon-number states is 
sub-Poissonian.    
Let us return to Eq.~(\ref{1.04}) and assume that
\begin{equation}
\hat \varrho_{\rm in}(n_{0}) = \hat \varrho_{{\rm in}1}
\otimes \hat \varrho_{{\rm in}2},
\label{AD4}
\end{equation}
where 
\begin{equation}
\hat \varrho_{{\rm in}2} = \sum_{n_{0}} \tilde p_{n_{0}} \,
|n_0\rangle\langle n_0| \, .
\label{AD0}
\end{equation}
To be more specific, let us consider (as an example of a sub-Poissonian
distribution) a binomial probability 
distribution,
\begin{equation}
\tilde p_{n_{0}}
= {N\choose n_0} p^{n_0}(1-p)^{N-n_0} \quad {\rm if} \quad n_{0}\leq N
\label{AD1}
\end{equation}
and $\tilde p_{n_{0}}$ $\!=$ $\!0$ elsewhere ($0$ $\!<$ $\!p$ $\!<$ $\!1$).
Note that for \mbox{$p$ $\!\to$ $\!0$}, \mbox{$N$ $\!\to$ $\!\infty$}, and 
$pN$ finite, the binomial distribution (\ref{AD1}) reduces to a Poisson
distribution, with $\bar{n}$ $\!=$ $\!pN$ being the mean photon number.
Using Eqs.~(\ref{AD4}) and (\ref{AD0}), in place of Eq.~(\ref{1.3}) we  
easily find that detection of no photons in one of the output channels of 
the beam splitter now yields the conditional (mixed photon-added) state
\begin{equation}
\hat{\varrho}_{{\rm out}1}(m_{2}=0)
= \sum_{n_{0}} \tilde p_{n_{0}}\,\hat{\varrho}_{{\rm out}1}(n_{0},m_{2}=0)
= \sum_{n_{0},\Phi} \tilde p_{n_{0}} \tilde p_{\Phi}| \, 
\Psi_{n_0} \rangle \big\langle \Psi_ {n_0}| 
\label{AD5}
\end{equation}
in the other output channel.
Accordingly, the probability of detecting the state is 
the average of $P(n_{0})$ given in Eq.~(\ref{M1}), 
\begin{eqnarray}
P = \sum_{n_0} \tilde p_{n_{0}} P(n_{0}).
\label{AD6}
\end{eqnarray}

\begin{figure}
\centering\epsfig{figure=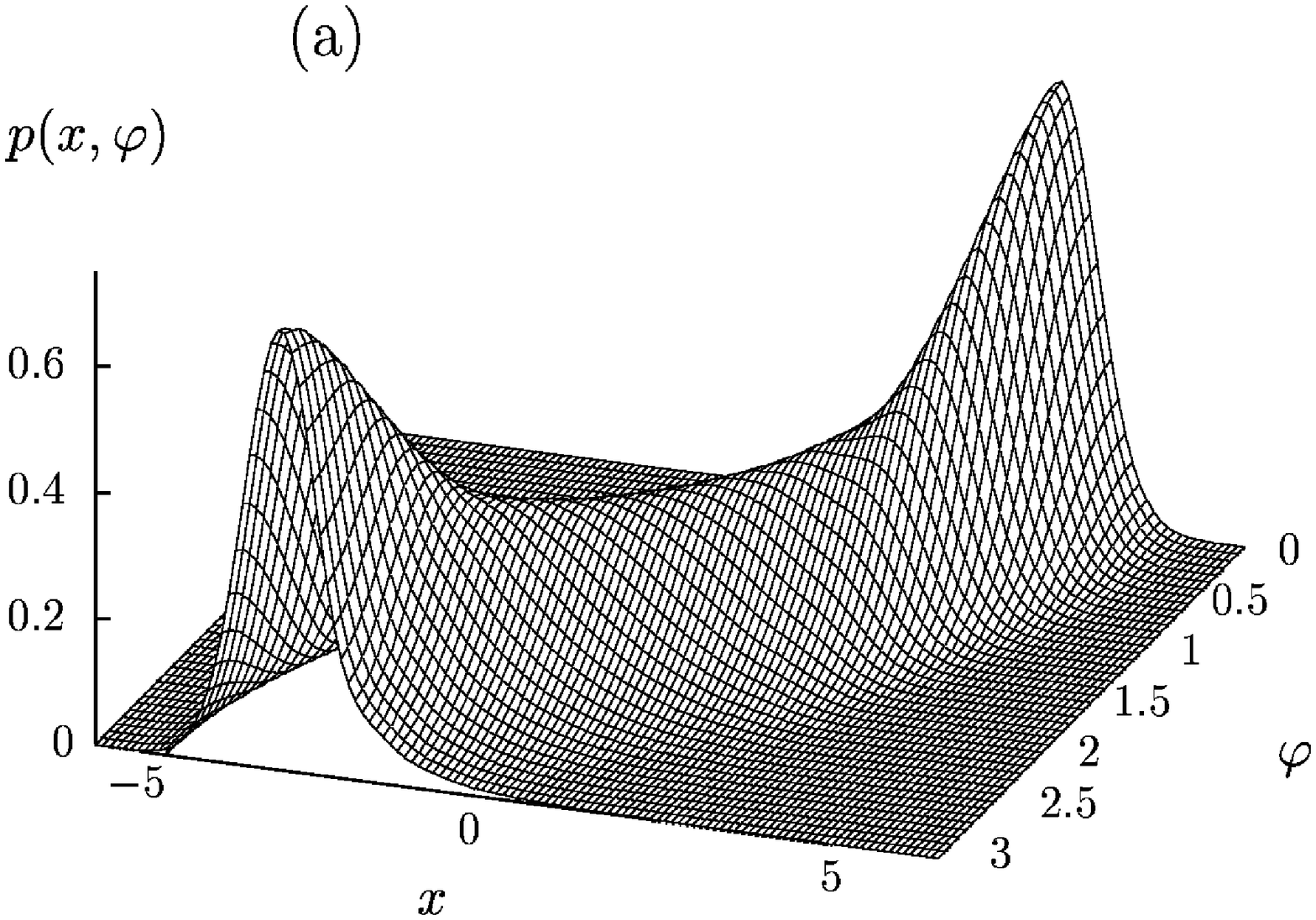,width=0.75\linewidth}

~
\vspace{0.1cm}

~

\centering\epsfig{figure=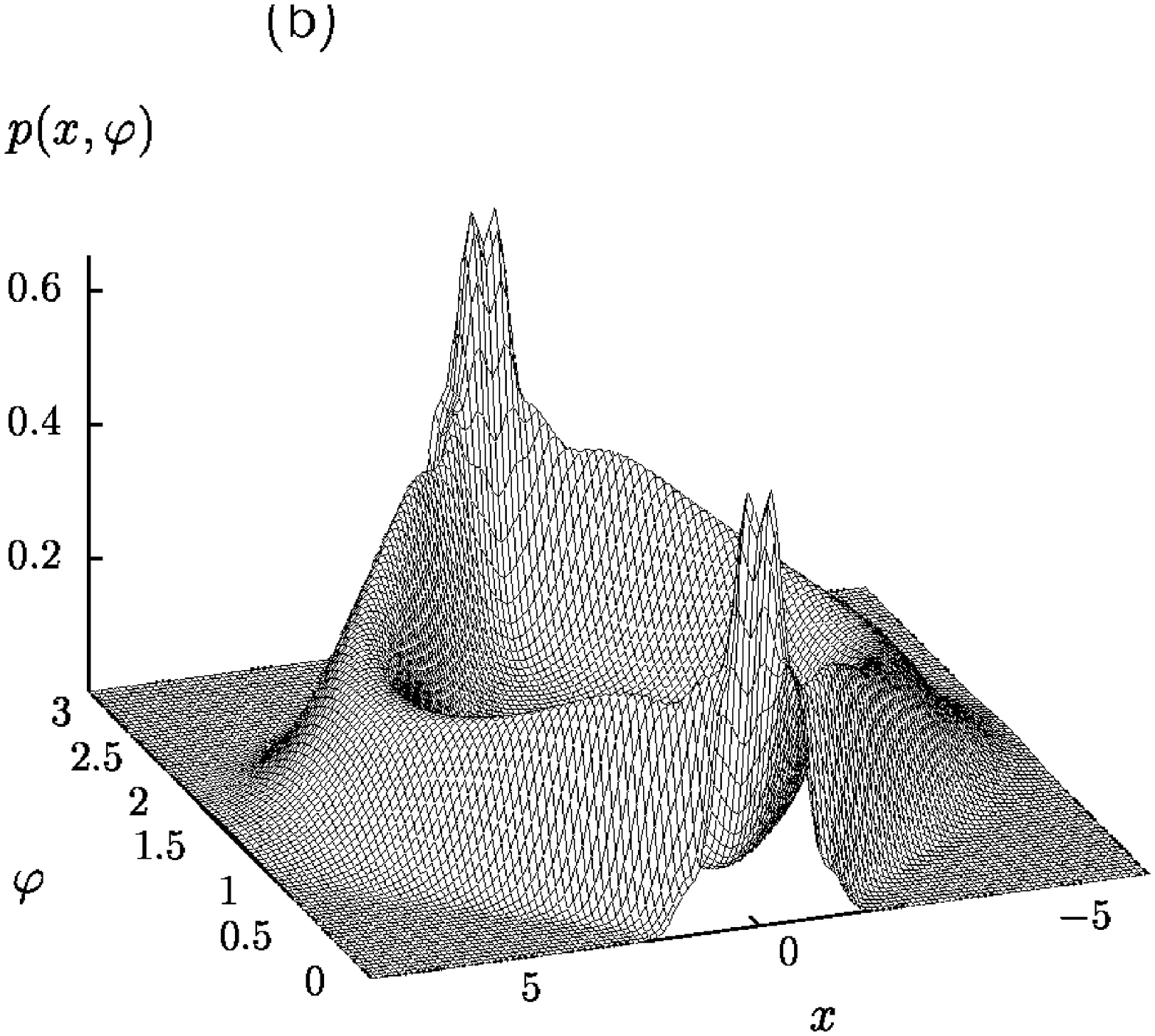,width=0.75\linewidth}
\caption{The quadrature-component distributions of mixed
photon-added coherent (a) and squeezed vacuum (b) states for 
$p$ $\!=$ $\!0.8$ and $N$ $\!=$ $\!5$ in Eq.~(\protect\ref{AD1}),
the values of the other parameters being the same as in 
Figs.~\protect\ref{Fig3} and \protect\ref{Fig6}.
\label{Fig9}}
\end{figure}

Examples of the quadrature-component distributions 
$p(x,\varphi)$ $\!=$ $\!\sum_{n_{0}}\tilde p_{n_{0}}\,p_{n_{0}}(x,\varphi)$
of mixed photon-added coherent and squeezed vacuum states 
are plotted in Figs.~\ref{Fig9}(a) 
and (b), respectively. In the figures it is assumed that 
\mbox{$p$ $\!=$ $\!0.8$} and $N$ $\!=$ $\!5$ [i.e., the mean photon number and 
the photon-number variance are $\bar{n}$ $\!=$ $\!4$ and 
$\overline{(\Delta n)^{2}}$ $\!=$ $\!0.8$, respectively].
Comparing Figs.~\ref{Fig9}(a) and (b) with Figs.~\ref{Fig3}(b)
and \ref{Fig6}(b), respectively, we see that the typical nonclassical 
features (such as squeezing and quantum interference) are preserved, 
even when the photon number state $|n_{0}\rangle$ is replaced with a 
sub-Poissonian mixed state (\ref{AD0}) (i.e., a smeared photon-number state).
As expected, the probabilities (\ref{AD6}) of observing the mixed 
photon-added states become smaller than those obtained for pure 
Fock-state inputs [$P\approx\!0.07\%$ for the mixed photon-added 
coherent state in Fig.~\ref{Fig9}(a) and $P\approx\!0.04\%$ for 
the mixed photon-added squeezed vacuum state in Fig.~\ref{Fig9}(b)]. 


\section{Conclusion}
\label{sec5}

We have studied the problem of generating photon-added states
using conditional output measurement on a beam splitter.
When a single-mode radiation field is mixed with a mode prepared 
in a photon-number number state, then the mode in one of the output 
channels photon-added is prepared in a photon-added state, provided
that in the other output channel no photons are detected. 
We have studied the conditions under which the photon-added
states can be regarded as photon-added input states or states that
belong the same class of states as the photon-added input states do. 
Typical examples of input states to which photons can be added in this way
are thermal states, coherent states, squeezed states, displaced 
photon-number states, and coherent phase states.
 
Photon-added states are highly nonclassical states in general.
In particular, photon-added coherent states can be regarded as 
non-Gaussian squeezed states, as can be seen, e.g., from the  
derived expression for the quadrature-component distribution.
Another interesting class of
states that can be produced in the way described are photon-added
squeezed vacuum states, which exhibit all the typical properties of 
Schr\"odinger-cat-like states. Photon adding to a squeezed vacuum
can therefore be regarded as a method for producing Schr\"odinger-cats.
We have analyzed the photon-added squeezed vacuum states in terms
of the photon-number and the quadrature-component distributions
and phase-space functions, such as the Wigner and Husimi functions,
and have presented both analytical and numerical results. We have
shown that photon-added squeezed vacuum states may be
regarded as superpositions of two non-Gaussian squeezed coherent
states that tend to the familiar (Gaussian) squeezed
coherent states as the number of added photons goes to infinity. 

With regard to possible experimental implementations, we
have also performed calculations allowing for an input mode
prepared in a statistical mixture of Fock states in place of
a pure Fock state. As expected, mixtures of Fock states give rise to 
conditional states that can be regarded, in a sense,
as smeared photon-added states. It is worth noting that when the 
photon-number distribution of the mixtures is typically sub-Poissonian, 
then the characteristic nonclassical features of photon-added states 
can be observed. In the paper we have demonstrated the effect of smearing
assuming a binomial photon-number distribution. In particular, 
combining mixed Fock states of that type with a squeezed vacuum,
the conditional output states also exhibit, in general, properties that
are typically observed for Schr\"{o}dinger-cat-like states. That
is to say, the characteristic properties of squeezed vacuum states
are (apart of some smearing) preserved. Clearly, when the photon-number 
distribution of the mixed Fock states becomes Poissonian, then the 
nonclassical properties that are typically associated with photon-added 
states disappear.


\section*{Acknowledgment}
This work was supported by the Deutsche Forschungsgemeinschaft.
We would like to thank T. Opatrn\'{y} and E. Schmidt for 
valuable discussions.


\begin{appendix}
\renewcommand{\thesection}{Appendix \Alph{section}}
\section{Derivation of Eq.~(\protect\ref{M21})}
\label{app1}
\setcounter{equation}{0}
\renewcommand{\theequation}{\Alph{section} \arabic{equation}}
To find the asymtotoic form (\ref{M21}) of $Q_{n_0}^{(\pm)}(x,p)$
defined in Eq.~(\ref{M20}), we first consider the asymptotic behavior of 
$|\alpha|^{2n_0}e^{-|\alpha|^2}$ for $n_{0}$ $\!\to$ $\!\infty$,
\begin{eqnarray}
|\alpha|^{2n_0}e^{-|\alpha|^2} & \approx &
\!\frac{n_0!}{2\pi\!n_0}
\,\exp\!\left(-\frac{|\alpha^2-n_0|^2}{2n_0}\right)
\nonumber \\ & = & 
\frac{n_0!}{2\pi\!n_0}
\,\exp\!\left(-\frac{|\alpha + \sqrt{n_0}|^2\;
|\alpha-\sqrt{n_0}|^2} {2n_0}\right),
\label{A1}
\end{eqnarray}
which may be approximated by the sum of two Gaussians centered at 
$\alpha$ $\!=$ $\!\pm\sqrt{n_0}$, 
\begin{eqnarray}
|\alpha|^{2n_0}e^{-|\alpha|^2}\approx\frac{n_0!}{2\pi\!n_0}
\,\left[\exp\!\left(-|\alpha\!-\!\sqrt{n_0}|^2\right)
+ \exp\!\left(-|\alpha\!+\!\sqrt{n_0}|^2\right)\right].
\label{A2}
\end{eqnarray}
Next let us consider the infinite-series approximation 
of the complementary complex error function \cite{Abramowitz1},
\begin{eqnarray}
\lefteqn{
{\rm Erfc}(x+iy) \approx {\rm Erfc}(x)
+ \frac{e^{-x^2}}{2\pi x}\left[1-\cos(2xy)+i\sin(2xy)\right]
}
\nonumber \\ && \hspace{12ex} +\,
\frac{2e^{-x^2}}{\pi}\sum_{n=1}^{\infty}\frac{e^{-n^2/4}}{n^2+4x^2}
\left[f_n(x,y)+ig_n(x,y)\right],
\label{A3}
\end{eqnarray}
where $f_n(x,y)$ and $g_n(x,y)$ are given by
\begin{eqnarray}
\nonumber f_n(x,y)=2x[1-\cosh(ny)\cos(2x)]
+ n\sinh(ny)\sin(2xy)
\label{A4}
\end{eqnarray}
and
\begin{eqnarray}
g_n(x,y)=2x\cosh(ny)\sin(2x)
+ n\sinh(ny)\cos(2xy).
\label{A5}
\end{eqnarray}
  From Eq.~(\ref{A3}) [together with Eqs.~(\ref{A4}) and (\ref{A5})] 
we easily see that
\begin{equation}
{\rm Erfc(x+iy)} \approx {\rm Erfc(x)} \quad {\rm if} \quad |x|\gg 1.
\label{A6}
\end{equation}
Further the approximations   
\begin{eqnarray}
{\rm Erfc}(x) &\approx& 0 \quad {\rm if} \quad x \gg 1,
\label{A7} \\
{\rm Erfc}(-x)&\approx& 0 \quad {\rm if} \quad x \ll -1.
\label{A8}
\end{eqnarray}
are valid \cite{Abramowitz1}.
Recalling Eq.~(\ref{M20}), we see that the argument $x\!+\!iy$ in
Eq.~(\ref{A3}) corresponds to $\mp\sqrt{\textstyle\frac{1}{2}\kappa'}\alpha$.
Since for chosen $\kappa'$ ($\kappa'$ $\!\neq$ $\!0$) and sufficiently 
large $n_{0}$ the Gaussians in Eq.~(\ref{A2}) are nonzero only for large 
$|\alpha|$, we can apply the approximations (\ref{A6}) -- (\ref{A8}),
which [together with Eq.~(\ref{A2})] yields the asymptotic form (\ref{M21}).
Note that due to the error function only one of the two Gaussians in 
Eq.~(\ref{A2}) survives.
\end{appendix}


\newpage
\newpage
\newpage
\newpage
\newpage
\newpage
\newpage
\newpage
\newpage
\end{document}